\documentclass[final,5p,times,twocolumn]{elsarticle}

\usepackage{amssymb}

\usepackage{epsfig,amsmath,amssymb}
\usepackage{graphics} 
\usepackage{subfigure}
\usepackage{color}
\usepackage{mathrsfs}
\usepackage{natbib}
\usepackage{bbold}
\usepackage{bbm}
\biboptions{sort&compress}

\usepackage{wasysym}

\usepackage[
colorlinks=true,
linkcolor=blue,
urlcolor=blue,
citecolor=blue,
hyperindex=true
]{hyperref}

\journal{Journal of magnetism and magnetic materials}

\begin{document}

\begin{frontmatter}



\title{Frustrated spin-$\frac{1}{2}$ Heisenberg magnet on an $AA$-stacked honeycomb bilayer: High-order study of the collinear magnetic phases of the $J_{1}$--$J_{2}$--$J_{1}^{\perp}$ model}


\author{P H Y Li$^{1,2}$ and R F Bishop$^{1,2}$}
\ead{peggyhyli@gmail.com; raymond.bishop@manchester.ac.uk}

\address{$^{1}$ School of Physics and Astronomy, Schuster Building, The University of Manchester, Manchester, M13 9PL, UK}

\address{$^{2}$ School of Physics and Astronomy, University of Minnesota, 116 Church Street SE, Minneapolis, Minnesota 55455, USA}

\begin{abstract}
The zero-temperature phase diagram of the frustrated spin-$\frac{1}{2}$ $J_{1}$--$J_{2}$--$J_{1}^{\perp}$ Heisenberg magnet on an $AA$-stacked honeycomb bilayer lattice is studied using the coupled cluster method implemented to very high orders.  On each monolayer the spins interact via nearest-neighbor (NN) and frustrating next-nearest-neighbor isotropic antiferromagnetic Heisenberg interactions with respective strength parameters $J_{1}>0$ and $J_{2}\equiv\kappa J_{1}>0$.  The two layers are coupled such that NN interlayer pairs of spins also interact via a similar isotropic Heisenberg interaction of strength $J_{1}^{\perp}\equiv \delta J_{1}$, which may be of either sign.  In particular, we locate with high accuracy the complete phase boundaries in the $\kappa$-$\delta$ half-plane with $\kappa>0$ of the two quasiclassical collinear antiferromagnetic phases with N\'{e}el or N\'{e}el-II magnetic order in each monolayer, and the interlayer NN pairs of spins either aligned (for $\delta<0$) or anti-aligned (for $\delta > 0$) to one another.  Compared to the two-sublattice N\'{e}el order, in which all NN intralayer pairs of spins are antiparallel to one another, the four-sublattice N\'{e}el-II order is characterized by NN intralayer pairs of spins on the honeycomb lattice being antiparallel to one another along zigzag (or sawtooth) chains in a specified direction from among the three equivalent honeycomb-lattice directions, and parallel to one another for the corresponding interchain pairs.    
\end{abstract}


\end{frontmatter}


\section{Introduction}
\label{introd_sec}
Frustrated quantum spin-lattice models in two dimensions, in which a variety of different phases can emerge from zero-point quantum fluctuations even at zero temperature ($T=0$), are prototypical models in which to study quantum phase transitions as some system parameter is varied at $T=0$.  The possible phases obviously include quasiclassical states with definite magnetic long-range order (LRO), such that SU(2) spin-rotational symmetry is broken and the average local on-site magnetization (proportional to $\langle \mathbf{s}_{j} \rangle$, where $\mathbf{s}_{j}$ is the spin operator on lattice $j$) is nonzero.  More interestingly, however, they also include quantum paramagnetic phases with no classical counterparts, for which $\langle \mathbf{s}_{j} \rangle=0$, and hence all magnetic LRO has melted.

Such quantum paramagnetic phases that preserve the SU(2) spin-rotational symmetry may or may not still break other lattice symmetries.  The former case includes a variety of valence-bond crystalline (VBC) phases in which one or more lattice symmetries are broken by the formation of spin-singlet states involving static spin complexes arranged in some regular pattern.  For example, a plaquette VBC phase typically breaks only translational symmetry, while a dimer VBC phase breaks both translational and rotational lattice symmetries.  By contrast, quantum spin liquid (QSL) phases \cite{Savary:2017_QSL} conserve {\it all}\, lattice symmetries, and any other symmetries of the Hamiltonian, by definition.  QSL phases themselves may be either gapless or gapped, the latter typically with some topological order (e.g., of the $Z_{2}$ type).  They are archetypal disordered ground states that possess massive many-body entanglement.

A typical generic system in which strong quantum fluctuations can combine with spin frustration to produce magnetically disordered ground-state (GS) phases is the $J_{1}$--$J_{2}$ Heisenberg model on a given two dimensional (2D) lattice, in which both nearest-neighbor (NN) and next-nearest-neighbor (NNN) Heisenberg exchange interactions, of respective strengths $J_{1}>0$ and $J_{2} \equiv \kappa J_{1}>0$, compete with one another.  By now there have been very many calculations on a variety of 2D lattices (e.g., square, honeycomb, triangular, kagome, and other Archimedean lattices) that show that each of these $J_{1}$--$J_{2}$ spin-lattice systems can exhibit such quantum paramagnetic GS phases, typically in the vicinity of the value for the frustration parameter $\kappa$ that denotes maximal frustration in the classical limit $s \to \infty$, where $s$ is the spin quantum number of each of the lattice spins.

Generally, for a lattice of given dimensionality (here taken to be equal to 2), quantum fluctuations tend to be larger for lower values of both the spin quantum number $s$ and the lattice coordination number $z$.  For these reasons the spin-$\frac{1}{2}$ $J_{1}$--$J_{2}$ Heisenberg magnet on the honeycomb lattice occupies a special niche in the present context since it is the simplest (in the sense of being composed of only one type of polygon, i.e., the hexagon) of the four 2D Archimedean lattices (all comprised of arrangements of regular polygons, with every site equivalent to all others) that take the lowest coordination, $z=3$.  Each Archimedean lattice is uniquely defined by specifying the ordered sequence of polygons that surround each of the equivalent vertices.  The other three Archimedean lattices with $z=3$ all comprise lattices with more than one type of polygon.  For example, of the remaining two bipartite lattices with $z=3$, the CaVO ($4\cdot 8^2$) lattice comprises squares and octagons, while the SHD ($4\cdot 6\cdot 12$) lattice comprises squares, hexagons, and dodecagons.  The non-bipartite Archimedean lattice with $z=3$ is the star ($3\cdot 12^{2}$) lattice, which comprises triangles and dodecagons.  We note that of the four Archimedean lattices with lowest coordinations, $z=3$, the honeycomb ($6^{3}$) lattice is special in that it is the only one in which all of the edges are also equivalent.  The remaining three, namely the CaVO, SHD, and star lattices, all contain two different types of NN bonds.  

The honeycomb lattice is also unlike the other well-studied square, triangular, and kagome lattices for the spin-$\frac{1}{2}$ $J_{1}$--$J_{2}$ model in that it is non-Bravais.  Unlike these others it comprises two sites per unit cell, with the structure of two interlacing triangular Bravais sublattices.  An immediate consequence is that the spin-$\frac{1}{2}$ $J_{1}$--$J_{2}$ model on the honeycomb lattice is free from the restrictions, which apply to the corresponding model on Bravais lattices, that are imposed by the Lieb-Schultz-Mattis (LSM) theorem \cite{Lieb:1961_LSMH-theorem} and its generalizations due to Hastings \cite{Hastings:2004_Lieb-LSM-Hast-theorem} and others \cite{Affleck:1988_Lieb-LSM-Hast-theorem,Oshikawa:2000_Lieb-LSM-Hast-theorem,Watanabe:2015_Lieb-LSM-Hast-theorem}.  Broadly speaking, the extension by Hastings of the LSM theorem implies that any such spin-lattice system that has a half-odd-integer spin per unit cell cannot have both a unique ground state and a gap in the excitation spectrum.  Thus, for spin-$\frac{1}{2}$ quantum magnets on Bravais lattices, such as the square, triangular, and kagome lattices, a gapped spectrum strictly implies a degenerate GS phase.  This could be, for example, either a VBC phase caused by a lattice symmetry breaking, or a QSL phase (e.g., of the $Z_{2}$ variety) caused by topological degeneracy.  By contrast, for spin-$\frac{1}{2}$ models on the honeycomb lattice, with its two sites per unit cell, in principle one can have a gapped quantum paramagnetic QSL GS phase that does not break any symmetry.

For all of the above reasons the spin-$\frac{1}{2}$ $J_{1}$--$J_{2}$ model on the honeycomb-lattice monolayer has received an enormous amount of attention in recent years, with a large number of theoretical techniques being applied to it \cite{Rastelli:1979_honey,Mattsson:1994_honey,Fouet:2001_honey,Mulder:2010_honey,Okumura:2010_honey,Wang:2010_honey,Cabra:2011_honey,Ganesh:2011_honey_merge,Ganesh:2011_honey_errata_merge,Clark:2011_honey,DJJF:2011_honeycomb,Reuther:2011_honey,Albuquerque:2011_honey,Mosadeq:2011_honey,Oitmaa:2011_honey,Mezzacapo:2012_honey,Bishop:2012_honeyJ1-J2,Li:2012_honey_full,RFB:2013_hcomb_SDVBC,Zhang:2013_honey,Ganesh:2013_honey_J1J2mod_PRB87,Ganesh:2013_honey_J1J2mod-XXX,Zhu:2013_honey_J1J2mod-XXZ,Gong:2013_J1J2mod-XXX,Yu:2014_honey_J1J2mod,Ciolo:2014_honey_XY,Jiang:2016_SqLatt-honey,Ferrari:2017_honey_J1J2mod}.  Despite this intense activity, there is still no overall consensus on the structure of its ($T=0$) quantum phase diagram, particularly in the paramagnetic regime, as a function of the frustration parameter, $\kappa \equiv J_{2}/J_{1}$, as we discuss in more detail in Sec.\ \ref{model_sec}.  While almost all studies concur that the system retains N\'{e}el magnetic LRO for sufficiently weak frustration, $\kappa < \kappa_{c_{1}}^{>}$, there remain disagreements both over the nature of the stable GS phases for $\kappa > \kappa_{c_{1}}^{>}$ and the precise numerical value of the critical parameter $\kappa_{c_{1}}^{>}$ at which N\'{e}el order melts, although in the latter regard most recent calculations using theoretical techniques of high potential accuracy do give values for $\kappa_{c_{1}}^{>}$ in the approximate range $0.19 \lesssim \kappa_{c_{1}}^{>} \lesssim 0.23$.

If we restrict further discussion about these uncertainties to potentially high-accuracy methods performed either at high orders in some well-defined sequence of approximations or in large-scale numerical implementations, there is no doubt that they are in large part due to the twin facts that almost all such methods are biased in favor of some predetermined GS phase and/or are not performed from the outset in the infinite-lattice limit in which we are interested.  In the latter regard, for example, such potentially high-accuracy techniques as the exact diagonalization (ED) of finite-sized lattices comprising $N$ spins and the density-matrix renormalization group (DMRG) method always require some form of finite-size scaling to extrapolate to the thermodynamic limit, $N \to \infty$.  Such extrapolations have been shown explicitly (see, e.g., Ref.\ \cite{Sandvik:2012_SqLatt_J-Q_model}) to be capable of containing large uncertainties.  While this is particularly true in cases where theoretical considerations provide little or no information on which rigorously to base the extrapolation scheme, it can also even be the case when some such guidance exists.

Within this context the coupled cluster method (CCM) \cite{Coester:1958_ccm,Coester:1960_ccm,Cizek:1966_ccm,Kummel:1978_ccm,Bishop:1978_ccm,Bishop:1982_ccm,Arponen:1983_ccm,Bishop:1987_ccm,Arponen:1987_ccm,Arponen:1987_ccm_2,Bartlett:1989_ccm,Arponen:1991_ccm,Bishop:1991_TheorChimActa_QMBT,Bishop:1998_QMBT_coll,Zeng:1998_SqLatt_TrianLatt,Fa:2004_QM-coll,Bartlett:2007_ccm,Bishop:2014_honey_XXZ_nmp14,Farnell:2019_ccm_non-coplanar-mod-states} has come to occupy a special role in recent years since it is one of the very few high-accuracy methods of modern quantum many-body theory that is applied from the very outset in the thermodynamic limit, $N \to \infty$, and hence for which any need for finite-size scaling is always obviated.  It is in large part for that reason why it is the method we utilize in the present work.  Furthermore, it is now widely accepted that the CCM offers one of the most flexible, most widely applicable, and most accurate at a given level of computational resource, of all {\it ab initio}\, techniques that are available for dealing with a diverse range of problems in microscopic quantum many-body theory.  In particular, the CCM has already been applied to many different spin-lattice systems in the broad arena of quantum magnetism (see e.g., Refs.\ \cite{DJJF:2011_honeycomb,Bishop:2012_honeyJ1-J2,Li:2012_honey_full,RFB:2013_hcomb_SDVBC,Zeng:1998_SqLatt_TrianLatt,Fa:2004_QM-coll,Farnell:2019_ccm_non-coplanar-mod-states,Bishop:2014_honey_XY,Li:2014_honey_XXZ,Bishop:2015_honey_low-E-param,Bishop:2016_honey_grtSpins,Li:2016_honeyJ1-J2_s1,Bishop:2019_SqLatt_bilayer} and references contained therein), including both the spin-$\frac{1}{2}$ $J_{1}$--$J_{2}$ model on the honeycomb lattice \cite{Bishop:2012_honeyJ1-J2,RFB:2013_hcomb_SDVBC} of interest here, as well as its various extensions, e.g., to the spin-1 counterpart \cite{Li:2016_honeyJ1-J2_s1}, the case of anisotropic ($XXZ$) couplings \cite{Li:2014_honey_XXZ}, and the isotropic $XY$ version of the model \cite{Bishop:2014_honey_XY}.

Given the still unresolved nature of the $T=0$ quantum phase diagram of the spin-$\frac{1}{2}$ $J_{1}$--$J_{2}$ model on the honeycomb-lattice monolayer, it seems worthwhile to examine larger classes of models, to which they reduce as a special case.  One such direction that has already received much attention (see, e.g., Refs.\ \cite{Albuquerque:2011_honey,Li:2012_honey_full} and references contained therein) is to remain with the spin-$\frac{1}{2}$ honeycomb-lattice monolayer, but also now to include next-next-nearest-neighbor Heisenberg interactions of strength $J_{3}$, resulting in the so-called $J_{1}$--$J_{2}$--$J_{3}$ model.  Another, potentially even more revealing, extension is to the corresponding spin-$\frac{1}{2}$ $J_{1}$--$J_{2}$--$J_{1}^{\perp}$ model on a honeycomb bilayer, which is the model of interest here.  Each layer comprises an identical frustrated $J_{1}$--$J_{2}$ system, and the two layers are now connected by NN Heisenberg exchange bonds of strength $J_{1}^{\perp} \equiv \delta J_{1}$, with the layers arranged in $AA$ stacking [i.e., with each site of one (horizontal) monolayer placed vertically above its equivalent on the other].

\begin{figure*}[t]
\begin{center}
\mbox{
\subfigure[]{\includegraphics[width=2.8cm]{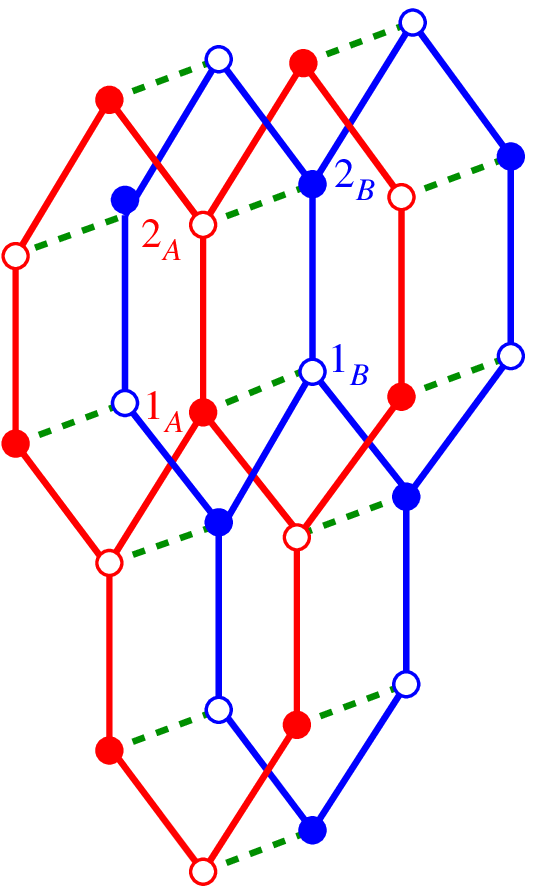}}
\hspace{0.5cm}
\subfigure[]{\includegraphics[width=3.7cm]{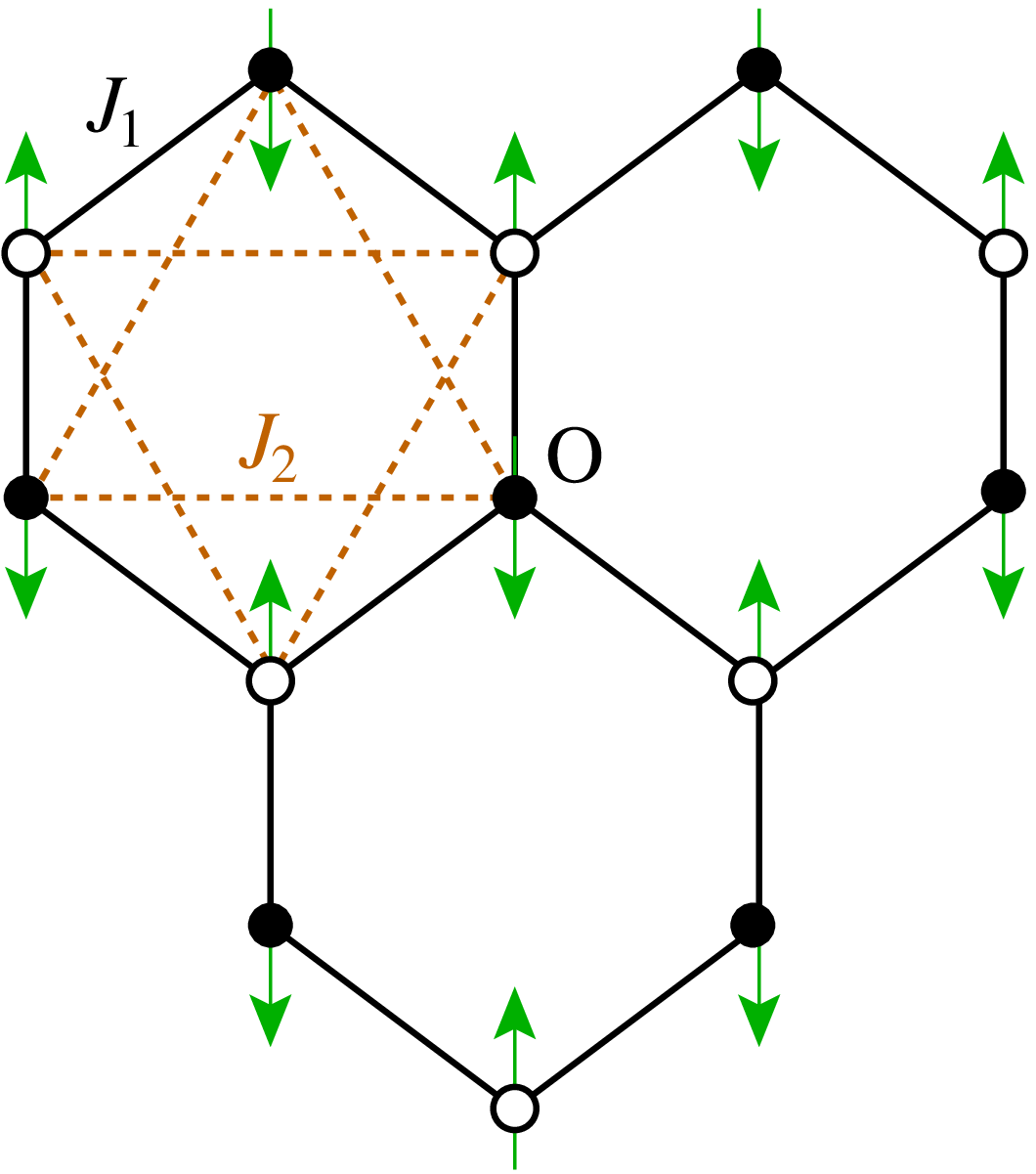}}
\hspace{0.5cm}
\subfigure[]{\includegraphics[width=3.7cm]{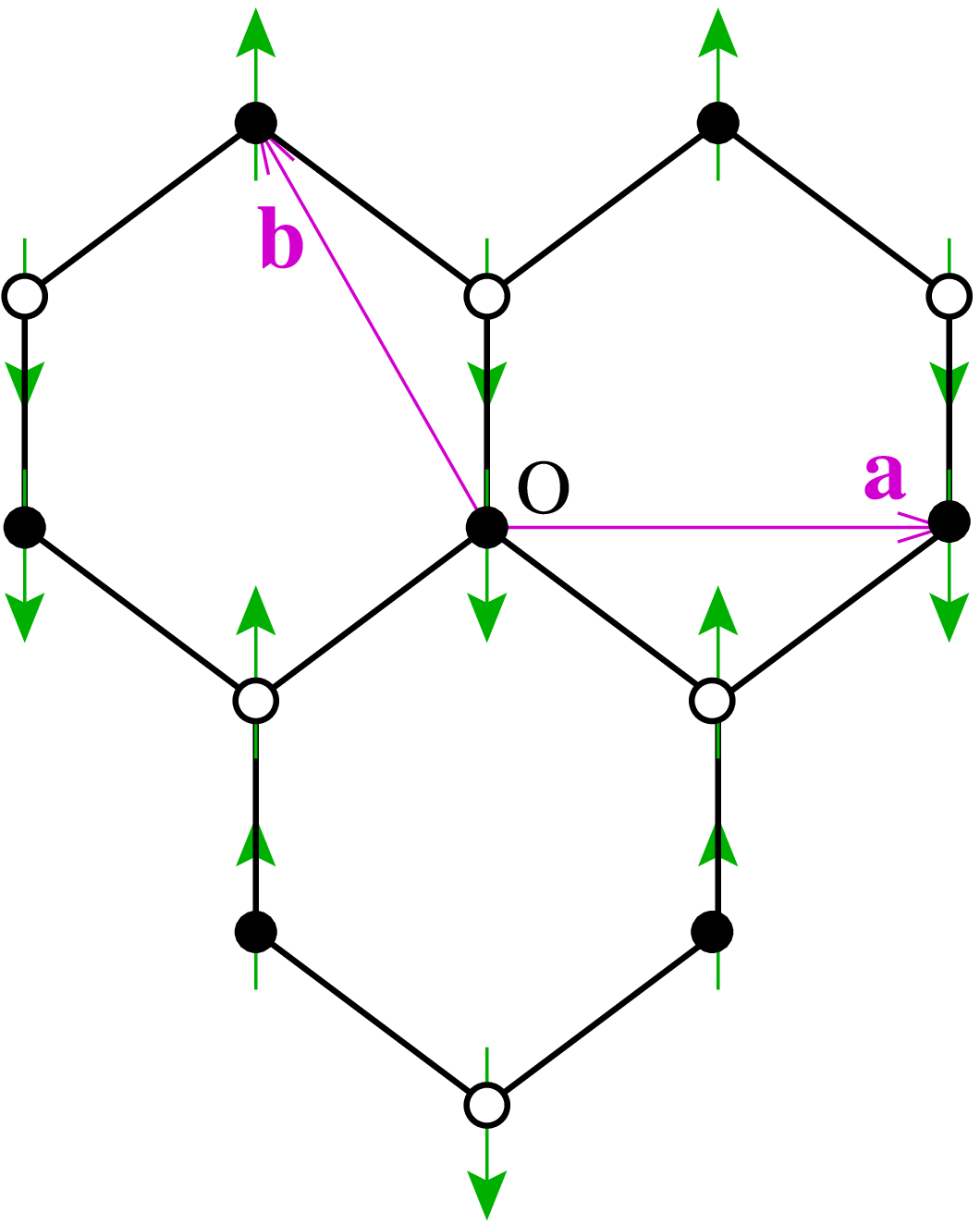}} 
\hspace{0.5cm}
\subfigure{\includegraphics[width=2.5cm]{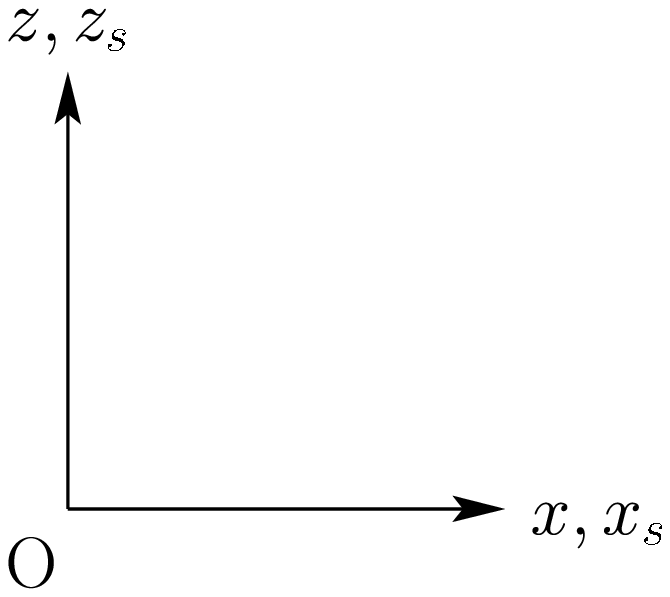}}
}
  \caption{The $J_{1}$--$J_{2}$--$J_{1}^{\perp}$ model on the bilayer honeycomb 
    lattice, showing (a) the two layers $A$ (red) and $B$ (blue), the nearest-neighbor (NN) bonds ($J_{1} = $ -----; $J_{1}^{\perp} = $ - - -)  and the four sites ($1_{A}, 2_{A}, 1_{B}, 2_{B}$) of the unit cell; (b) the intralayer bonds ($J_{1} = $ -----; $J_{2} = $ - - -) on each monolayer and the monolayer N\'{e}el state; and (c) the triangular Bravais lattice vectors ${\bf a}$ and ${\bf b}$, and one of the three equivalent monolayer N\'{e}el-II states.  Sites ($1_{A}, 2_{B}$) and ($2_{A}, 1_{B}$) on the two triangular lattices of each monolayer are shown by filled and empty circles respectively, and the spins are represented by the (green) arrows on the lattice sites.  For both states on the bilayer, spins on NN sites between the two layers are antiparallel (parallel) for $\delta>0$ $(\delta<0)$.}
  \label{model_pattern}
 \end{center}
\end{figure*}

The $J_{1}$--$J_{2}$--$J_{1}^{\perp}$ model is particularly interesting since the $J_{2}$ and $J_{1}^{\perp}$ bonds act in quite different ways to destroy the N\'{e}el magnetic LRO that exists when the (antiferromagnetic) NN $J_{1}\; (> 0)$ bonds prevail.  Thus, while the inclusion of (antiferromagnetic) NNN $J_{2}\; (> 0)$ bonds acts to frustrate the NN $J_{1}$ bonds by tending to favor another form of magnetic LRO, the NN $J_{1}^{\perp}$ bonds of either sign {\it do not\,} directly frustrate the N\'{e}el order.  However, they {\it do\,} compete with the $J_{1}$ bonds by promoting interlayer NN dimers to develop, which are spin-singlet (spin-triplet) pairs when $J_{1}^{\perp} > 0\; (J_{1}^{\perp} < 0)$.  The inclusion of interlayer $J_{1}^{\perp}$ bonds thus promotes a competition with the intralayer $J_{1}$ bonds between a phase with N\'{e}el magnetic LRO on each monolayer and a nonclassical paramagnetic phase of the dimerized VBC variety, wherein the (independent) dimers are now between NN interlayer pairs.  This effect is purely quantum-mechanical in origin, since in the classical ($s \to \infty$) limit the $J_{1}^{\perp}$ bonds have no effect at all on the intralayer N\'{e}el ordering.  {\it A priori}, the effect is  expected to be maximal for the case $s=\frac{1}{2}$ considered here.

While the spin-$\frac{1}{2}$ $J_{1}$--$J_{2}$--$J_{1}^{\perp}$ model on a honeycomb bilayer (with $J_{2} \equiv \kappa J_{1}; J_{1}^{\perp} \equiv \delta J_{1}; J_{1}>0$) has received some attention in the last few years, almost all of the effort has focused on calculating the boundary of the N\'{e}el phase in the quadrant of the $\kappa$-$\delta$ plane with $\kappa >0$, $\delta>0$ \cite{Zhang:2014_honey_bilayer,Arlego:2014_honey_bilayer,Bishop:2017_honeycomb_bilayer_J1J2J1perp,Li:2017_honeycomb_bilayer_J1J2J1perp_Ext-M-Field}.  Nevertheless, the spin-$\frac{1}{2}$ $J_{1}$--$J_{2}$ model on a honeycomb monolayer has also been shown to have another form of magnetic LRO in the range $\kappa_{c_{2}}^{<} < \kappa < \kappa_{c_{2}}^{>}$ , where $\kappa_{c_{2}}^{<} > \kappa_{c_{1}}^{>}$, which is not present in the classical ($s \to \infty$) version of the model except precisely at the point $\kappa = \frac{1}{2}$ where it is degenerate with a class of spiral states.  This phase, the so-called N\'{e}el-II phase, discussed more fully in Sec.\ \ref{model_sec}, and which corresponds to phase IV in Ref.\ \cite{Fouet:2001_honey}, is actually the $T=0$ GS phase in the classical $J_{1}$--$J_{2}$--$J_{3}$ honeycomb-lattice model only for values $J_{3}<0$ (when $J_{1}>0$) \cite{Rastelli:1979_honey,Fouet:2001_honey}, where, although it is then actually degenerate with an infinite manifold of noncoplanar spin configurations \cite{Fouet:2001_honey}, it is then stabilized by quantum fluctuations \cite{Fouet:2001_honey} via the order-by-disorder mechanism \cite{Villain:1977_ordByDisord_merge,Villain:1980_ordByDisord_merge}.

It has been shown by several authors \cite{Albuquerque:2011_honey,Li:2012_honey_full} that for the spin-$\frac{1}{2}$ case of the honeycomb-lattice $J_{1}$--$J_{2}$--$J_{3}$ model the N\'{e}el-II phase is actually stabilized over a finite region of the phase space with $J_{i}>0$, for $i=1,2,3$ (and see, in particular, Fig.\ 2 in each of Refs.\ \cite{Albuquerque:2011_honey,Li:2012_honey_full}).  For this reason it is of considerable interest to investigate also the phase boundary of the N\'{e}el-II phase for the spin-$\frac{1}{2}$ $J_{1}$--$J_{2}$--$J_{1}^{\perp}$ model on the honeycomb bilayer, which to our knowledge, has only been previously investigated \cite{Li:2019_honeycomb_bilayer_J1J2J1perp_neel-II} in the half-plane $\delta > 0$, as also for the N\'{e}el phase.  Again, to the best of our knowledge, there have been no prior investigations of the stability of both of these collinear magnetic orderings (i.e., N\'{e}el and N\'{e}el-II) on each monolayer of the bilayer model, in the case of ferromagnetic (FM) interlayer coupling ($\delta < 0$), despite the fact that this case is particularly interesting in its own right, for reasons that we now elaborate.

Thus, in the limit $\delta \to -\infty$, the system clearly simulates exactly a spin-1 $J_{1}$--$J_{2}$ model on a honeycomb-lattice monolayer.  While far fewer theoretical studies exist for this model than for its spin-$\frac{1}{2}$ counterpart, two recent calculations, one using the density-matrix renormalization group (DMRG) on cylindrical systems \cite{Gong:2015_honey_J1J2mod_s1} and another using the CCM employed here \cite{Li:2016_honeyJ1-J2_s1} both agree that each of the quasiclassical N\'{e}el and N\'{e}el-II antiferromagnetic (AFM) phases also form stable GS phases in the spin-1 honeycomb-lattice monolayer version of the $J_{1}$--$J_{2}$ model, just as in the spin-$\frac{1}{2}$ case discussed above.  Thus, we can use the spin-$\frac{1}{2}$ $J_{1}$--$J_{2}$--$J_{1}^{\perp}$ bilayer model as a tool to interpolate, as a function of the interlayer coupling $\delta$, between the spin-$\frac{1}{2}$ ($\delta=0$) and spin-1 ($\delta=-\infty$) $J_{1}$--$J_{2}$ honeycomb-lattice monolayer models, with the possibility in so doing of shedding more light on the collinear AFM phases of both models.

\section{The model}
\label{model_sec}
The $J_{1}$--$J_{2}$--$J_{1}^{\perp}$ model on a honeycomb-lattice
bilayer, shown in Fig.\ \ref{H_eq}, has a Hamiltonian given by
\begin{equation}
\begin{aligned}
H & =  J_{1}\sum_{{\langle i,j \rangle},\alpha} \mathbf{s}_{i,\alpha}\cdot\mathbf{s}_{j,\alpha} + 
J_{2}\sum_{{\langle\langle i,k \rangle\rangle},\alpha} \mathbf{s}_{i,\alpha}\cdot\mathbf{s}_{k,\alpha}  \\
& \quad + J_{1}^{\perp}\sum_{i} \mathbf{s}_{i,A}\cdot\mathbf{s}_{i,B} \\
& \equiv  J_{1}h(\kappa,\delta)\,; \quad  \kappa \equiv J_{2}/J_{1}\,, \quad \delta \equiv J_{1}^{\perp}/J_{1}\,,
\label{H_eq}
\end{aligned}
\end{equation}
where the sums over $\langle i,j \rangle$ and $\langle\langle i,k \rangle\rangle$ run, respectively, over all NN and NNN pairs of spins in each (horizontal) monolayer, counting each pair once and once only, and where the layer index $\alpha$ runs over the two monolayers, labelled $A$ and $B$.  The two monolayers are arranged in $AA$ stacking, such that each spin $\mathbf{s}_{i,A}$ on layer $A$ lies vertically above its counterpart $\mathbf{s}_{i,B}$ on layer $B$.  The sites on each monolayer are arranged in a regular honeycomb-lattice pattern, and each site $s_{i,\alpha}$ carries a spin-$s$ particle described by the usual SU(2) spin operators $\mathbf{
  s}_{i,\alpha}\equiv(s^{x}_{i,\alpha},s^{y}_{i,\alpha},s^{z}_{i,\alpha})$, with $\mathbf{s}_{i,\alpha}^{2} = s(s+1)\mathbbm{1}$.

We shall restrict attention here to the extreme quantum case, $s=\frac{1}{2}$.  We further restrict ourselves to the most interesting case where both intralayer Heisenberg bonds (i.e., NN bonds and NNN bonds) are AFM in nature (i.e., $J_{1}>0$ and $J_{2} \equiv \kappa J_{1} > 0$), such that they tend to frustrate one another.  The parameter $J_{1}$ then merely acts to set the overall energy scale, and the Hamiltonian may thus be written as $H=J_{1}h(\kappa,\delta)$, as in the last line of Eq.\ (\ref{H_eq}), thereby explicitly demonstrating that the relevant model parameters are only $\kappa$ and $\delta$.  Our aim is thus to examine the $T=0$ quantum phase diagram of the spin-$\frac{1}{2}$ version of the $J_{1}$--$J_{2}$--$J_{1}^{\perp}$ model Hamiltonian of Eq.\ (\ref{H_eq}) on a honeycomb bilayer in the $\kappa$-$\delta$ half-plane with $\kappa>0$.  Although we also concentrate here on the hitherto unexplored regime of FM interlayer coupling ($\delta < 0$), we shall also present some results with $\delta > 0$ for completeness.

Our primary interest will be to investigate the regimes of stability in the $\kappa$-$\delta$ plane of the two collinear AFM phases (i.e., the quasiclassical N\'{e}el and N\'{e}el-II phases) on each monolayer as the interlayer coupling parameter $\delta$ is introduced.  The $AA$-stacked bilayer honeycomb-lattice itself is shown in Fig.\ \ref{model_pattern}(a), and the spin arrangements of the N\'{e}el and N\'{e}el-II phases on each monolayer are illustrated respectively in Figs.\ \ref{model_pattern}(b) and \ref{model_pattern}(c).  Clearly, the N\'{e}el state has AFM sawtooth (or zigzag) chains (i.e., with spins that alternate in direction along the chains) in each of the three equivalent honeycomb-lattice directions.  The N\'{e}el-II state, by contrast has such AFM sawtooth chains along only one of the three directions, as shown in Fig.\ \ref{model_pattern}(c), for example, along the $x$ direction.  The NN spins between adjacent such AFM sawtooth chains in the N\'{e}el-II state are aligned parallel to one another.  The pattern of such adjacent parallel spin pairs in the N\'{e}el-II state is identical to that of the dimers in the staggered-dimer VBC (SDVBC) state, such that the N\'{e}el-II and SDVBC state break the same lattice symmetries (i.e., rotational and translational).  For this reason many theoretical investigations find it particularly difficult to differentiate these two states in the $T=0$ quantum phase diagram of any such model that supports one or other of them (e.g., Ref.\ \cite{Albuquerque:2011_honey}).  The SDVBC itself is also sometimes known as the lattice nematic state (see, e.g., Ref.\ \cite{Mulder:2010_honey}).

\section{Methodology}
\label{ccm_sec}
Since the CCM formalism has been discussed extensively elsewhere in the literature (see, e.g., Refs.\ \cite{Bishop:1991_TheorChimActa_QMBT,Bishop:1998_QMBT_coll,Fa:2004_QM-coll} and references contained therein), we confine ourselves here to a brief discussion of its most important features for the problem at hand.  The exact ket and bra GS wave functions are defined to be $|\Psi\rangle$ and $\langle\tilde{\Psi}| \equiv \langle\Psi|/\langle\Psi|\Psi\rangle$.  The system under study has a Hilbert space that we assume may be described in terms of a normalised model (or reference) state $|\Phi\rangle$, $\langle\Phi|\Phi\rangle = 1$, which acts as a cyclic vector for a corresponding set of {\it mutually commuting\,} multiconfigurational creation operators $\{C_{I}^{+}\}$, $[C_{I}^{+}, C_{J}^{+}]=0$, such that $|\Phi\rangle$ is a generalized vacuum state with respect to them, $C_{I}^{-}|\Phi\rangle = 0 = \langle\Phi|C_{I}^{+}\,,\forall I \neq 0$, in a notation where we define $C_{I}^{-} \equiv (C_{I}^{+})^{\dag}$ and $C_{0}^{+} \equiv \mathbbm{1}$.  The GS ket $|\Psi\rangle$ is chosen within the CCM to have the intermediate normalization, $\langle\Phi|\Psi\rangle = 1$.

The index $I$ is a set index and, in general, $C_{I}^{+}$ comprises a product of single-particle operators, as we describe more fully below in the current context of spin-lattice models.  The set $\{I\}$ is complete in the usual sense that the set of states $\{C_{I}^{+}|\Phi\rangle\}$ provides a complete basis for the ket Hilbert space.  Although not vital, it is also convenient to choose the basis to be orthonormalized, such that $\langle\Phi|C_{I}^{-}C_{J}^{+}|\Phi\rangle = \delta_{IJ}$, where $\delta_{IJ}$ is an appropriately defined Kronecker symbol.

The GS wave functions $|\Psi\rangle$ and $\langle\tilde{\Psi}|$ are now formally parametrized in the CCM, in a independent fashion, in terms of the distinctive exponentiated forms involving the correlation operators,

\begin{equation}
  |\Psi\rangle = e^{S}|\Phi\rangle\,, \quad \langle\tilde{\Psi}| = \langle\Phi|\tilde{S}e^{-S}\,;    \label{wave_functn_Psi_Eq}
\end{equation}

\begin{equation}
  S=\sum_{I \neq 0}{\cal S}_{I}C_{I}^{+}\,, \quad \tilde{S}=\mathbbm{1} + \sum_{I \neq 0}\tilde{\cal S}_{I}C_{I}^{-}\,,  \label{sum_create_destruct_operators_Eq}
  \end{equation}
  which are a hallmark of the method.  Since $S$ and $\tilde{S}$ are henceforth treated as independent operators, the Hermiticity relation, $\langle\tilde{\Psi}|=(|\Psi\rangle)^{\dag}/\langle\Psi|\Psi\rangle$, may be violated when subsequent approximations are made, typically by truncating the complete set of configurations $\{I\}$ in the sums of Eq.\ (\ref{sum_create_destruct_operators_Eq}), as described more fully below.  This possible shortcoming is far outweighed in practice by the fact that the CCM parametrizations of Eqs.\ (\ref{wave_functn_Psi_Eq}) and (\ref{sum_create_destruct_operators_Eq}) always {\it exactly\,} satisfy the very important Hellmann-Feynman theorem at {\it all\,} such levels of approximation \cite{Bishop:1998_QMBT_coll}.  It is precisely this feature that guarantees the robustness and accuracy of numerical results obtained within the CCM framework.

  The GS expectation value $\bar{X}$ of an arbitrary operator $X$ may thus be expressed within the CCM purely in terms of the $c$-number correlation coefficients $\{{\cal S}_{I}, \tilde{{\cal S}}_{I}\}$ as
  \begin{equation}
    \bar{X}=\bar{X}({\cal S}_{I}, \tilde{{\cal S}}_{I}) \equiv \langle\Phi|\tilde{S}e^{-S}Xe^{S}|\Phi\rangle\,.  \label{bar-X-Eq}
  \end{equation}
  The coefficients $\{{\cal S}_{I}, \tilde{{\cal S}_{I}}\}$ are themselves obtained, by minimizing the GS energy functional, $\bar{H}=\bar{H}({\cal S}_{I}, \tilde{{\cal S}}_{I})$, as obtained from Eq.\ (\ref{bar-X-Eq}) by replacing the arbitrary operator $X$ by the Hamiltonian $H$, with respect to all members of the set with $I \neq 0$ separately.  The GS expectation value $\bar{X}$ (and $\bar{H}$ in particular) is evaluated in practice by employing the nested commutation expansion for the CCM similarity transform $e^{-S}Xe^{S}$ of the operator $X$,
  \begin{equation}
    e^{-S}Xe^{S} = \sum_{n=0}^{\infty}\frac{1}{n!}[X, S]_{n}\,,  \label{similarity transform_eq}
  \end{equation}
in terms of the $n$-fold nested commutators $[X, S]_{n}$, which are defined iteratively as follows,
  \begin{equation}
    [X, S]_{n} \equiv [[X, S]_{n-1}, S], \quad [X, S]_{0} = X\,.
    \end{equation}

    The CCM parametrizations of Eqs.\ (\ref{wave_functn_Psi_Eq}) and (\ref{sum_create_destruct_operators_Eq}) are specifically chosen so that the otherwise infinite sum in Eq.\ (\ref{similarity transform_eq}) will actually terminate {\it exactly\,} at a low finite order for any operator $X$ all of whose terms involve only a product of a finite number of single-particle operators, such as the present Hamiltonian of Eq.\ (\ref{H_eq}).  The reason for this termination lies simply in the twin facts that all components of $S$ in the expansion of Eq.\ (\ref{sum_create_destruct_operators_Eq}) commute with one another, and the basic single-particle operators form a Lie algebra, together with the model state $|\Phi\rangle$ being a vacuum state for all operators $C_{I}^{-}$.  This same key feature of the CCM is also responsible for all terms in the expansion of $\bar{H}$ from Eq.\ (\ref{bar-X-Eq}) being linked.  It is precisely this fact that implies immediately that the CCM automatically satisfies exactly the Goldstone linked-cluster theorem at any level of truncation in the expansions of Eq.\ (\ref{sum_create_destruct_operators_Eq}) for the correlation operators.  In turn this also immediately guarantees the size-extensivity of the method at all such levels of approximation.  Thus, in the solution of the explicit CCM equations for the basic amplitudes $\{{\cal S}_{I}, \tilde{{\cal S}}_{I}\}$ that follow from the extremization of $\bar{H}$ (viz., $\delta\bar{H}/\delta\tilde{{\cal S}}_{I} = 0 = \delta\bar{H}/\delta{\cal S}_{I}, \forall I \neq 0$), the only approximation that is ever made in any CCM calculation is to decide which configurations $\{I\}$ are to be retained in the expansion of Eq.\ (\ref{sum_create_destruct_operators_Eq}), as described for the problem at hand below.

    A simple choice of CCM model state $|\Phi\rangle$ for any spin-lattice problem is any quasiclassical state with perfect magnetic LRO.  All such states are specified uniquely by fixing the spin on each lattice site independently in terms of its projection onto some given spin quantization axis.  For the present application we will thus use both of the N\'{e}el and N\'{e}el-II states on each honeycomb-lattice monolayer as our CCM model states.  In order to treat each lattice site and each spin for {\it any\,} such quasiclassical model state $|\Phi\rangle$ as being completely equivalent to one another it is now useful to choose a local spin quantization axis (or, equivalently, by making a suitable passive rotation in spin space) on every lattice site independently so that in such local axes the model state is one in which every spin points in the same downwards direction (i.e., along the negative $z$ direction), $|\Phi\rangle = |\!\downarrow\downarrow\downarrow\cdots\downarrow\rangle$.  Such a description has the effect that all such spin-lattice calculations may, from this point on, be treated on an equal footing and via a universal computational code.  All that then distinguishes one case from another is the resulting spin Hamiltonian, which hence needs to be re-expressed in terms of the local spin axes now specified separately for each model state.

    The magnetic order parameter is now taken to be the sublattice magnetization or average local on-site magnetization, $M$.  In the {\it local\,} rotated spin axes defined above, this simply takes the universal form
    \begin{equation}
      M=-\frac{1}{N}\sum_{i=1}^{N}\langle\Phi|\tilde{S}e^{-S}s_{l_{i}}^{z}e^{S}|\Phi\rangle\,,
    \end{equation}
    where the index $l_{i}\equiv(k_{i},\alpha)$ labels the sites on both bilayers $\alpha=A,B$ of the system, and $N(\rightarrow\infty)$ is the total number of spins on the bilayer.

    The above choice of local spin axes now ensures that the multiconfigurational creation operators $C_{I}^{+}$ may be chosen to be products of single spin-raising operators $s_{l_{i}}^{+}\equiv s_{l_{i}}^{x}+is_{l_{i}}^{y}$.  The set indices $I$ hence simply become sets of bilayer lattice site indices, $\{{I}\} \rightarrow \{{l_{1}, l_{2}, \cdots, l_{n}; n = 1, 2, \cdots, 2sN}\}$, in which any given site index $l_{i}$ may be repeated, but so that it appears at most $2s$ times in the most general case where each site carries a spin-$s$ particle.  Thus, we have $C_{I}^{+} \rightarrow s_{l_{1}}^{+}s_{l_{2}}^{+}\cdots s_{l_{n}}^{+}$, with $n=1, 2, \cdots, 2sN$.  For the present paper we restrict ourselves to the case $s=\frac{1}{2}$, for which quantum effects are expected to be the largest, so that no lattice site in any multiconfigurational index $I$ may appear more than once.

    As we have already intimated above, the {\it only\,} approximation that we now ever make is to restrict the configurations $\{{I}\}$ that we retain in the expansions of Eq.\ (\ref{sum_create_destruct_operators_Eq}) for the CCM correlation operators, $S$ and $\tilde{S}$.  For the present calculations we shall utilize the well-tested and widely used localized (lattice-animal-based subsystem) LSUB$n$ hierarchy of approximations.  At a specified $n$th order in the LSUB$n$ scheme one retains all multispin-flip correlations corresponding to all possible configurations of spins on the lattice that are confined to clusters of at most $n$ contiguous sites.  A set $I$ of sites is defined to be contiguous in this sense if every site of the set is a NN to at least one other site of the set, in some specified geometry.

    In order to reduce the number of {\it independent\,} multispin-flip configurations at a given LSUB$n$ order to the minumum, $N_{f}=N_{f}(n)$, at a given LSUB$n$ order, we utilize all of the space- and point-group symmetries of both the system Hamiltonian and the particular CCM model state being used.  In the same vein we also employ any relevant conservation laws.  For example, both the quasiclassical N\'{e}el and N\'{e}el-II AFM states are eigenstates of the operator $s_{T}^{z}\equiv\sum_{k,\alpha}s_{k,\alpha}^{z}$, using {\it global\,} spin axes, which represents the total $z$ component of spin for the system as a whole.  Both states have corresponding eigenvalue $s_{T}^{z}=0$, and accordingly all of the multispin-flip configurations $I$ retained in the expansions of the CCM correlation operators $S$ and $\tilde{S}$ given by Eq.\ (\ref{sum_create_destruct_operators_Eq}) are chosen to follow this conservation law.

    Even after all such symmetries and conservation laws have been fully incorporated the minimum number of independent LSUB$n$ configurations, $N_{f}(n)$, grows rapidly with the truncation index $n$, typically exponentially as $n$ becomes large.  The available computer power then typically determines the maximum order $n$ that can be calculated in practice for a given model.  In the present case, for the spin-$\frac{1}{2}$ $J_{1}$--$J_{2}$--$J_{1}^{\perp}$ model on the $AA$-stacked honeycomb bilayer lattice, by employing both massive parallelization and large-scale supercomputing resources, as well as a purpose-built and customized computer algebra package \cite{ccm_code} to both derive and solve \cite{Zeng:1998_SqLatt_TrianLatt} the large sets of $N_{f}$ coupled CCM equations for the GS ket- and bra-state coefficients $\{{{\cal S}_{I},\tilde{{\cal S}}_{I}}\}$, we can perform LSUB$n$ calculations up to the very high order $n=10$.  Thus, for the model under consideration, when the two monolayers are coupled ferromagnetically (i.e., when $\delta<0$), we have $N_{f}(10)=64\,780\; (183\,939)$ for the GS properties when the CCM model state is chosen so that each monolayer has N\'{e}el (N\'{e}el-II) AFM order.  The corresponding numbers for the case when the two monolayers are coupled antiferromagnetically (i.e., when $\delta>0$) are $N_{f}(10)=70\,118\; (197\,756)$.

\begin{figure*}[t]
\begin{center}
\mbox{
\subfigure[]{\includegraphics[width=5.5cm]{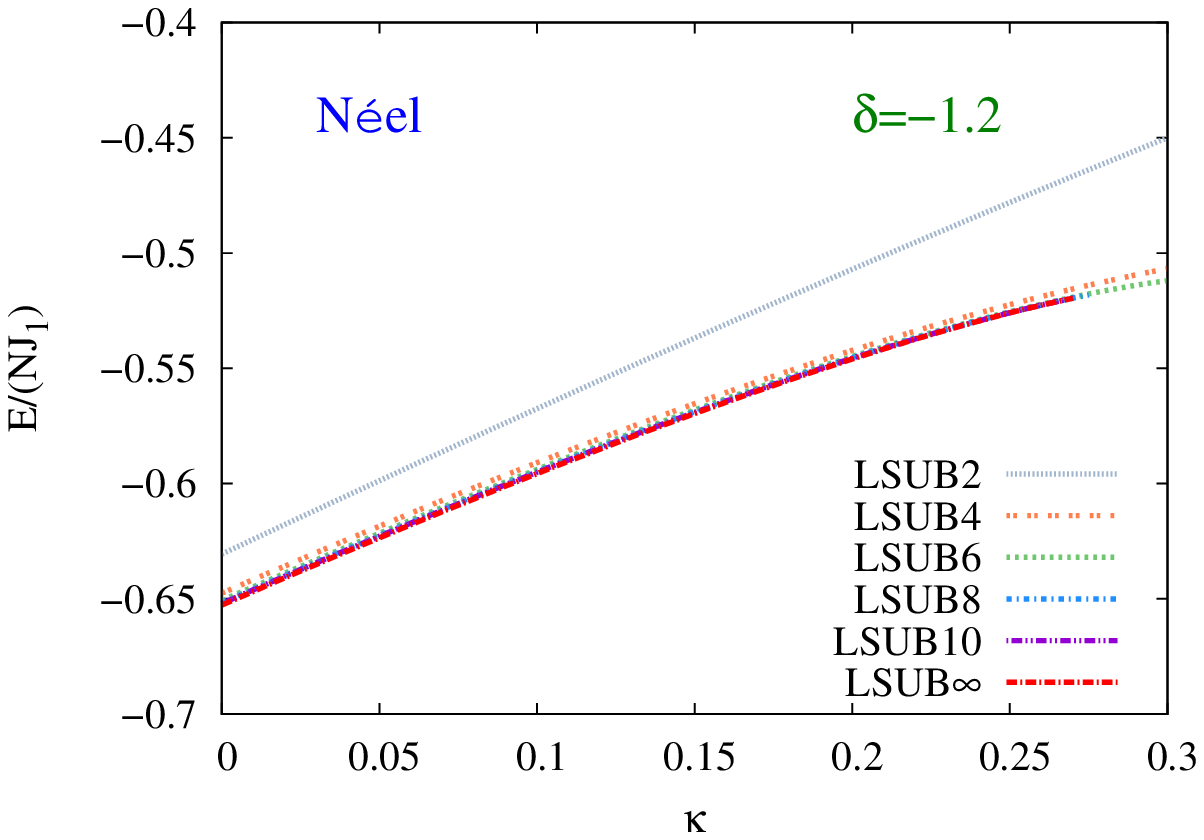}}
\hspace{0.1cm} \subfigure[]{\includegraphics[width=5.5cm]{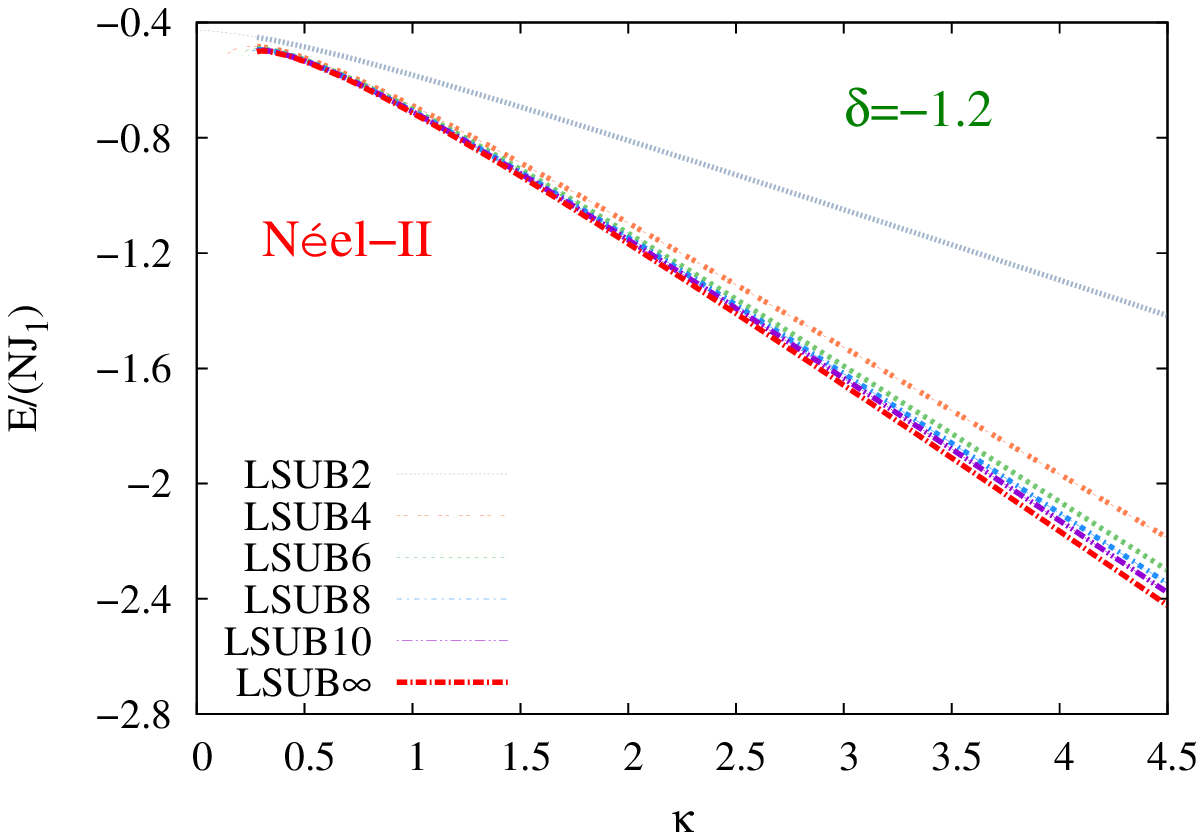}}
\hspace{0.1cm} \subfigure[]{\includegraphics[width=5.5cm]{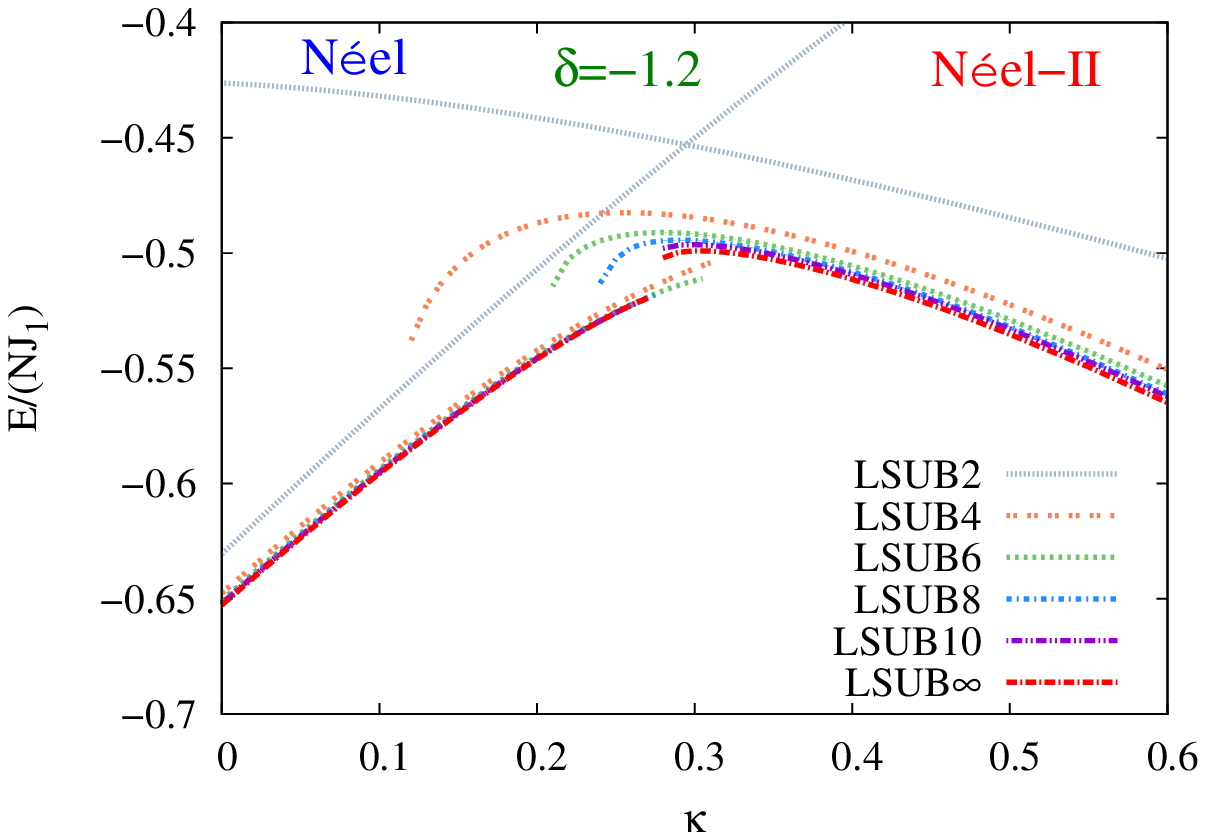}}
}
\caption{CCM results for the GS energy per spin $E/N$ (in units of $J_{1}$) versus the
  intralayer frustration parameter, $\kappa \equiv J_{2}/J_{1}$, for
  the spin-$\frac{1}{2}$ $J_{1}$--$J_{2}$--$J_{1}^{\perp}$ model on
  the bilayer honeycomb lattice (with $J_{1}>0$), for a selected value
  ($\delta=-1.2$) of the scaled interlayer exchange coupling constant,
  $\delta \equiv J_{1}^{\perp}/J_{1}$.  Results based on (a) the
  N\'{e}el state and (b) the N\'{e}el-II state on each monolayer, and
  the two layers coupled so that NN spins between them are parallel to
  one another, as CCM model states are shown in LSUB$n$ approximations
  with $n=2,4,6,8,10$, together with the corresponding LSUB$\infty$
  extrapolated result based on Eq.\ (\ref{extrapo_E}) and
  the LSUB$n$ data sets $n=\{2,6,10\}$.  Subfigure (c) shows an exploded view of the intermediate region.}
\label{E_neel_neel-II_fix-J1perp}
\end{center}
\end{figure*}      

It is clear that our LSUB$n$ approximation becomes asymptotically exact as the truncation index grows, $n \to \infty$.  Since no approximations are ever made in calculating any given LSUB$n$ approximant for any GS observable, it then follows that the only source of errors in our CCM calculations is in the very last step where the sequences of LSUB$n$ approximants are themselves extrapolated to the exact, $n \to \infty$, limit.  A great deal of empirical evidence now exists on how to perform such extrapolations, based on a large corpus of applications to diverse spin-lattice systems.  For example, a well tested and highly accurate extrapolation scheme for the GS energy per spin, $E/N$, and see, e.g., Refs.\ \cite{Bishop:2000_XXZ,Kruger:2000_JJprime,Fa:2001_SqLatt_s1,Fa:2004_QM-coll,Darradi:2005_Shastry-Sutherland,Bi:2008_EPL_J1J1primeJ2_s1,Bi:2008_JPCM_J1xxzJ2xxz_s1,Bi:2009_SqTriangle,Bishop:2010_UJack,Bishop:2010_KagomeSq,Bishop:2011_UJack_GrtSpins,DJJF:2011_honeycomb,PHYLi:2012_honeycomb_J1neg,Bishop:2012_honeyJ1-J2,Bishop:2012_honey_circle-phase,Li:2012_honey_full,PHYLi:2012_SqTriangle_grtSpins,Li:2012_anisotropic_kagomeSq,RFB:2013_hcomb_SDVBC,DJJFarnell:2014_archimedeanLatt,Bishop:2015_honey_low-E-param,Bishop:2016_honey_grtSpins,Li:2016_honey_grtSpins,Li:2016_honeyJ1-J2_s1}) is given by 
\begin{equation}
\frac{E(n)}{N} = e_{0}+e_{1}n^{-2}+e_{2}n^{-4}\,.     \label{extrapo_E}
\end{equation}    
By fitting our LSUB$n$ approximants $E(n)/N$ to Eq.\ (\ref{extrapo_E}) we may thereby obtain the extrapolated (exact LSUB$\infty$) value $e_{0}$.

  The GS expectation values of other physical observables typically converge slower than the energy, as is wholly to be expected.  Thus, for example, the LSUB$n$ approximants $M(n)$ to the magnetic order parameter of strongly frustrated systems, particularly for those with a quantum phase transition between states with and without magnetic LRO, such as our present model, have been found (and see, e.g., Refs.\ \cite{Bi:2008_JPCM_J1J1primeJ2,Bi:2008_PRB_J1xxzJ2xxz,Darradi:2008_J1J2mod,Reuther:2011_J1J2J3mod,Gotze:2012,Bishop:2014_honey_XXZ_nmp14,Bi:2008_EPL_J1J1primeJ2_s1,Bi:2008_JPCM_J1xxzJ2xxz_s1,Li:2019_honeycomb_bilayer_J1J2J1perp_neel-II}) to be accurately fitted by the well tested extrapolation scheme
\begin{equation}
M(n) = \mu_{0}+\mu_{1}n^{-1/2}+\mu_{2}n^{-3/2}\,.   \label{M_extrapo_frustrated}
\end{equation}
Once again, the extrapolated (exact LSUB$\infty$) value $\mu_{0}$ for $M$ may then be obtained by fitting a sequence of LSUB$n$ approximants $M(n)$ to Eq.\ (\ref{M_extrapo_frustrated}).

When making use of extrapolation schemes such as those in Eqs.\ (\ref{extrapo_E}) and (\ref{M_extrapo_frustrated}) one needs to be aware of any ``staggering effects'' that may be present in the approximant sequences.  For example, it is well known in perturbation theory that there usually exists an odd/even or $(2m-1)/2m$ (where $m \in \mathbb{Z}^{+}$ is a positive integer) staggering effect in the sequence of $n$th order approximants for various physical observables.  In such cases, where we often know {\it exact\,} extrapolation schemes, both the $n=(2m-1)$ and $n=2m$ subsequences obey an extrapolation scheme of the same sort (i.e., with identical exponents in the leading and sub-leading terms), but where the respective coefficients of the terms other than that corresponding to the extrapolated $(n \to \infty)$ value itself may differ.  In such cases one should clearly not mix odd- and even-order terms together in a single extrapolation sequence unless the staggering is separately incorporated properly.  In general it is difficult to include the staggering explicitly in a robust fashion, and one then usually separately extrapolates the odd terms and the even terms.  Our CCM SUB$n$ sequences of approximants for all physical observables also display such an odd/even staggering to a greater or lesser degree, dependant on both the model and the observable.  Since the Hamiltonian of Eq.\ (\ref{H_eq}) is itself bilinear in the single-spin operators, it is then natural to confine ourselves to extrapolating the even-order $(n=2m)$ approximants for any physical observable in any CCM spin-lattice calculation for the model.

While the above $(2m-1)/2m$ staggering in LSUB$n$ sequences of approximants for any physical quantity is common to all spin-lattice models on all lattices, honeycomb lattices tend to exhibit an {\it additional\,} $(4m-2)/4m$ staggering in the even subsequences, as has been noted elsewhere \cite{Bishop:2012_honeyJ1-J2,RFB:2013_hcomb_SDVBC,Li:2016_honeyJ1-J2_s1,Li:2019_honeycomb_bilayer_J1J2J1perp_neel-II,Bishop:2017_honeycomb_bilayer_J1J2J3J1perp}, and which now seems to originate in the non-Bravais nature of the honeycomb lattice \cite{Li:2019_honeycomb_bilayer_J1J2J1perp_neel-II}, which itself comprises two interlacing triangular Bravais sublattices.  Each of these displays the above-mentioned $(2m-1)/2m$ staggering, and the composite honeycomb lattice then magnifies the effect twofold into the observed $(4m-2)/4m$ staggering of the even $(n=2m)$ subsequence and, presumably, also a $(4m-3)/(4m-1)$ staggering of the odd $(n=2m-1)$ subsequence.  Since, as we have noted above, we are restricted for the present model to LSUB$n$ approximants with $n\leq 10$, in order to take all of these staggering effects into account, we restrict all extrapolations such as those based on Eqs.\ (\ref{extrapo_E}) and (\ref{M_extrapo_frustrated}) to LSUB$n$ data sets with $n=\{2,6,10\}$ for the results that we present in Sec.\ \ref{results_sec}. 
  
\section{Results}
\label{results_sec}
\begin{figure*}[t]
\begin{center}
\mbox{
\subfigure[]{\includegraphics[width=8.5cm]{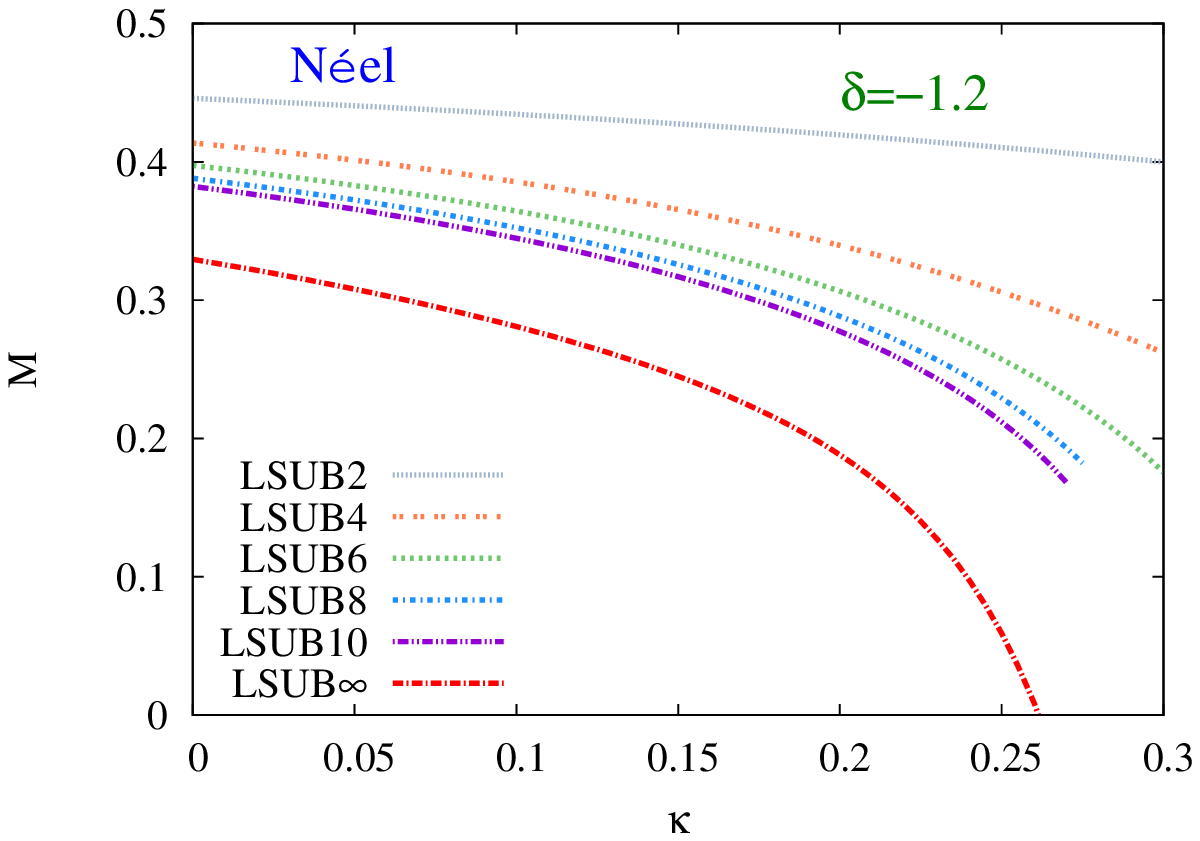}}
\hspace{0.1cm} \subfigure[]{\includegraphics[width=8.5cm]{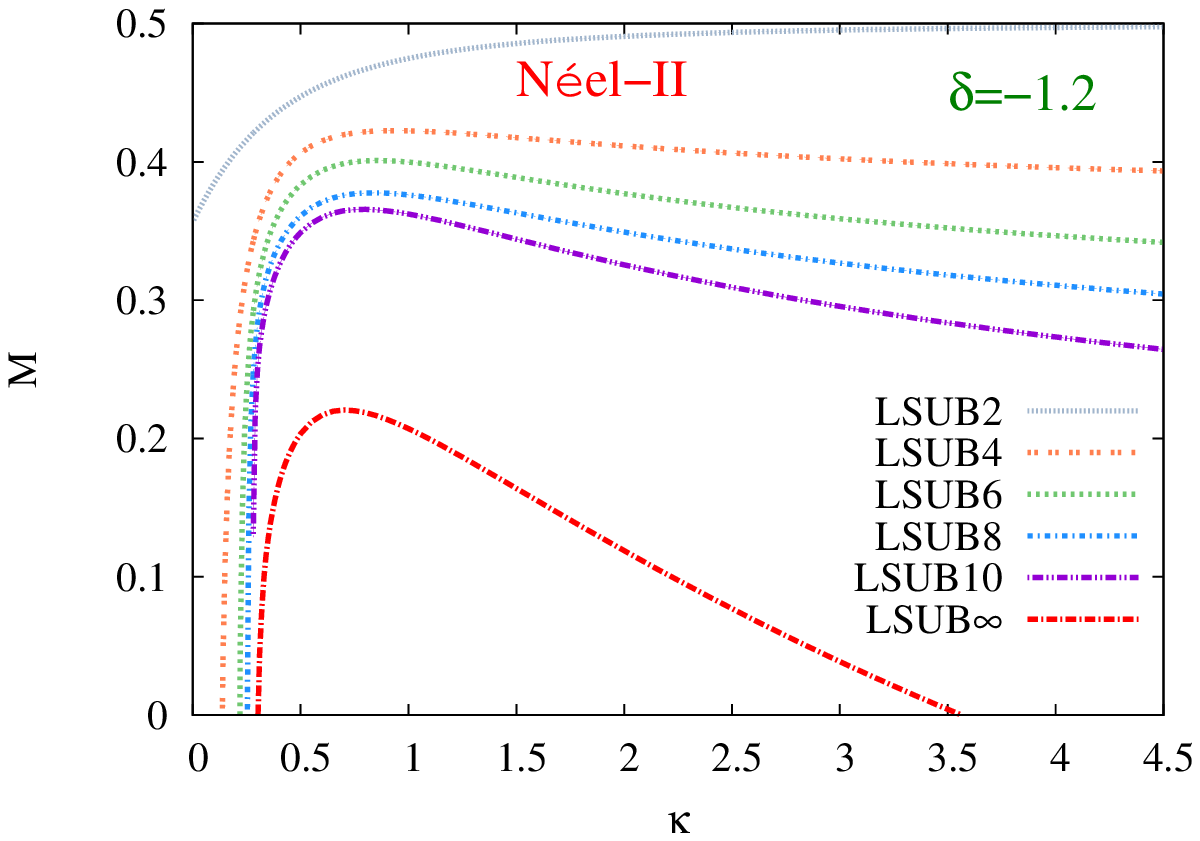}}
}
\caption{CCM results for the GS magnetic order parameter $M$ versus the
  intralayer frustration parameter, $\kappa \equiv J_{2}/J_{1}$, for
  the spin-$\frac{1}{2}$ $J_{1}$--$J_{2}$--$J_{1}^{\perp}$ model on
  the bilayer honeycomb lattice (with $J_{1}>0$), for a selected value
  ($\delta=-1.2$) of the scaled interlayer exchange coupling constant,
  $\delta \equiv J_{1}^{\perp}/J_{1}$.  Results based on (a) the
  N\'{e}el state and (b) the N\'{e}el-II state on each monolayer, and
  the two layers coupled so that NN spins between them are parallel to
  one another, as CCM model states are shown in LSUB$n$ approximations
  with $n=2,4,6,8,10$, together with the corresponding LSUB$\infty$
  extrapolated result based on Eq.\ (\ref{M_extrapo_frustrated}) and
  the LSUB$n$ data sets $n=\{2,6,10\}$.}
\label{M_neel_neel-II_fix-J1perp}
\end{center}
\end{figure*}

We first present results for the GS energy per spin (in units of $J_{1}$, $E/(NJ_{1})$, for our bilayer system, based separately on each of the N\'{e}el and N\'{e}el-II states on each monolayer as our CCM model state, in the regime of FM interlayer coupling, so that in the overall model state the two layers are coupled such that NN interlayer spin pairs are parallel to one another.  We show in Fig.\ \ref{E_neel_neel-II_fix-J1perp} results at a typical FM value, $\delta=-1.2$, of the interlayer coupling parameter as a function of the intralayer frustration parameter $\kappa$.  One clearly observes the very rapid convergence of the CCM LSUB$n$ sequences in each case.  We also show the corresponding LSUB$\infty$ extrapolated results, $e_{0}/J_{1}$, which are obtained from Eq.\ (\ref{extrapo_E}) and the corresponding LSUB$n$ data sets with $n=\{2,6,10\}$ as input to circumvent the $(4m-2)/4m$ staggering effect discussed above and which we will discuss below in more detail in connection with the corresponding results for the magnetic order parameter $M$, where we will also consider the natural transition points of the LSUB$n$ results that can be observed, for example, in Fig.\ \ref{E_neel_neel-II_fix-J1perp}(c).  We note from Fig.\ \ref{E_neel_neel-II_fix-J1perp} the extremely rapid convergence of the set of LSUB$n$ approximants to the GS energy, as the truncation index $n$ is increased, for both phases shown.  The results presented in Fig.\ \ref{E_neel_neel-II_fix-J1perp}(c) also clearly foreshadow the existence of an intermediate GS (non-magnetic) phase between the N\'{e}el and N\'{e}el-II bilayer phases for this model.

We turn next to our results for the magnetic order parameter $M$.  Thus, we show in Figs.\ \ref{M_neel_neel-II_fix-J1perp}(a) and \ref{M_neel_neel-II_fix-J1perp}(b) results at the same value $\delta=-1.2$ of the interlayer coupling as those shown in Figs.\ \ref{E_neel_neel-II_fix-J1perp}(a) and \ref{E_neel_neel-II_fix-J1perp}(b), respectively, for the GS energy.
We note immediately that, as is true in all practical implementations of the CCM, the LSUB$n$ approximants $M(n)$ extend beyond the actual (LSUB$\infty$) transition point(s) for the phase in question, for all finite values of the truncation index $n$, out to some corresponding termination point(s), beyond which no real solution exists for the respective LSUB$n$ equations.  Furthermore, these LSUB$n$ termination points clearly appear to converge uniformly to the LSUB$\infty$ estimates of the corresponding quantum critical points (QCPs) as $n$ is increased, as has been observed many times before (and see, e.g., Refs.\ \cite{Bishop:2012_honeyJ1-J2,RFB:2013_hcomb_SDVBC,Fa:2004_QM-coll,PHYLi:2012_honeycomb_J1neg}).  The $(4m-2)/4m$ staggering effect is clearly observed in the LSUB$n$ results for $M(n)$, and is particularly marked for the N\'{e}el-II results shown in Fig.\ \ref{M_neel_neel-II_fix-J1perp}(b).  Thus, the corresponding LSUB$\infty$ results $\mu_{0}$ for $M$ shown in Fig.\ \ref{M_neel_neel-II_fix-J1perp} are obtained from fitting Eq.\ (\ref{M_extrapo_frustrated}) to the respective LSUB$n$ data sets with $n=\{2,6,10\}$ only.

We note from Fig.\ \ref{M_neel_neel-II_fix-J1perp} that at all orders in our LSUB$n$ expansions of the CCM cluster operators the magnetic order parameter $M$ is overestimated, but as the truncation index $n$ is increased $M$ decreases.  This is fully as expected, since as one moves to higher values of the truncation index $n$ one is including more correlations.  These correlations simply involve clusters of ``wrong'' spins with respect to the corresponding perfectly ordered (N\'{e}el or N\'{e}el-II) classical state.  Clearly, the effect of systematically incorporating more and more such clusters of ``wrong'' spins will thus lead naturally to a systematic {\it reduction} in the order parameter, precisely as is observed in Fig.\ \ref{M_neel_neel-II_fix-J1perp} for both ordered antiferromagnetic phases.

Results such as these shown in Fig.\ \ref{M_neel_neel-II_fix-J1perp}(a) clearly indicate the existence of an upper QCP, $\kappa^{>}_{c_{1}}=\kappa^{>}_{c_{1}}(\delta)$, beyond which the N\'{e}el phase ceases to be a stable GS phase.  Thus, for example, we see from Fig.\ \ref{M_neel_neel-II_fix-J1perp}(a) that $\kappa_{c_{1}}^{>}(\delta=-1.2) \approx 0.261$, which may be compared with the corresponding CCM result for the $J_{1}$--$J_{2}$ honeycomb-lattice monolayer, $\kappa_{c_{1}}^{>}(\delta=0)\approx 0.183$, using the same LSUB$n$ data set with $n=\{2,6,10\}$ as input to Eq.\ (\ref{M_extrapo_frustrated}) \cite{Bishop:2017_honeycomb_bilayer_J1J2J1perp}.  Similarly, results such as those shown in Fig.\ \ref{M_neel_neel-II_fix-J1perp}(b) also indicate both a lower QCP, $\kappa_{c_{2}}^{<} = \kappa_{c_{2}}^{<}(\delta)$, and an upper QCP, $\kappa^{>}_{c_{2}}=\kappa_{c_{2}}^{>}(\delta)$, such that the N\'{e}el-II phase is a stable GS phase only in the regime $\kappa_{c_{2}}^{<}(\delta) < \kappa < \kappa_{c_{2}}^{>}(\delta)$ for a given value of the scaled interlayer coupling parameter $\delta$.  Thus, from Fig.\ \ref{M_neel_neel-II_fix-J1perp}(b) we see that $\kappa^{<}_{c_{2}}(\delta=-1.2) \approx 0.302$ and $\kappa_{c_{2}}^{>}(\delta=-1.2) \approx 3.563$, which values may be compared with the corresponding CCM results for the $J_{1}$--$J_{2}$ honeycomb-lattice monolayer, $\kappa_{c_{2}}^{<}(\delta=0) \approx 0.449$ and $\kappa_{c_{2}}^{>} \approx 1.487$, again using the same LSUB$n$ data sets with $n=\{2,6,10\}$ as input to Eq.\ (\ref{M_extrapo_frustrated}) \cite{Li:2019_honeycomb_bilayer_J1J2J1perp_neel-II}.

\begin{figure}[!b]
  \includegraphics[width=9cm]{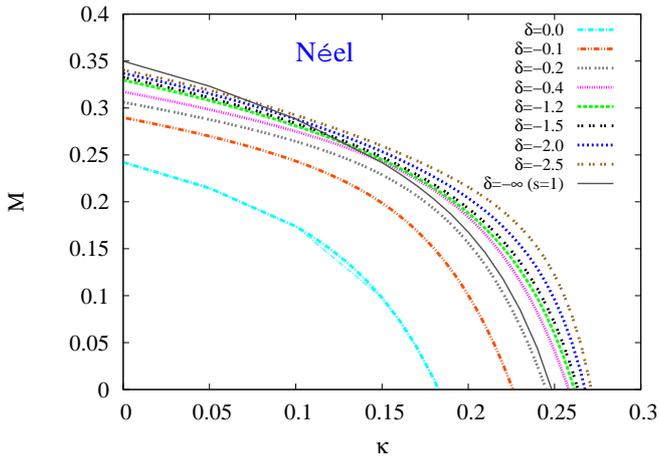}
  \caption{CCM results for the GS magnetic order parameter $M$ versus
    the intralayer frustration parameter, $\kappa \equiv J_{2}/J_{1}$,
    for the spin-$\frac{1}{2}$ $J_{1}$--$J_{2}$--$J_{1}^{\perp}$ model
    on the bilayer honeycomb lattice (with $J_{1}>0$), for a variety
    of ferromagnetic values of the scaled interlayer exchange coupling
    constant, $\delta \equiv J_{1}^{\perp}/J_{1}$.  In each case we
    show extrapolated results, based on the N\'{e}el state as CCM
    model state on each monolayer, and the two layers coupled so that
    NN spins between them are parallel to one another, obtained from
    using Eq.\ (\ref{M_extrapo_frustrated}) with the corresponding
    LSUB$n$ data sets $n=\{2,6,10\}$.  See the text for an explanation of the curve labelled $\delta=-\infty$ $(s=1)$.}
\label{M_multicontour_neel_honey_bilayer_AFM-FM}
\end{figure}

In Fig.\ \ref{M_multicontour_neel_honey_bilayer_AFM-FM} we show a set of LSUB$\infty$ extrapolated curves for the magnetic order parameter $M=M(\kappa)$ in the N\'{e}el GS phase, analogous to that shown in Fig.\ \ref{M_neel_neel-II_fix-J1perp}(a) for the specific value $\delta=-1.2$, but now for a variety of values of $\delta$ in the regime where the two layers are coupled ferromagnetically (so that NN interlayer pairs align parallel to one another), i.e., for $\delta < 0$. In each case we plot the value $\mu_{0}$ obtained from fitting  Eq.\ (\ref{M_extrapo_frustrated}) to the corresponding LSUB$n$ data for $M(n)$ with $n=\{2,6,10\}$.  We also show in Fig.\ \ref{M_multicontour_neel_honey_bilayer_AFM-FM} the corresponding CCM result for the N\'{e}el GS phase of the spin-1 $J_{1}$--$J_{2}$ model on a honeycomb-lattice monolayer \cite{Li:2016_honeyJ1-J2_s1}.  We note that in this case, instead of the LSUB$n$ approximation scheme, the alternative so-called SUB$n$-$n$ scheme has been used, as we now explain.

The CCM SUB$n$-$m$ approximation retains all multispin configurations $I$ in Eq.\ (\ref{sum_create_destruct_operators_Eq}) for the CCM correlation operators $S$ and $\tilde{S}$ that involve $n$ or fewer spin flips (with respect to the model state $|\Phi\rangle$) spanning a range of no more than $m$ contiguous sites.  In this context a single spin flip requires the action of the spin-raising operator $s^{+}_{k}$ acting once.  Clearly, for the spin-$\frac{1}{2}$ systems, the two truncation schemes LSUB$n$ and SUB$n$-$n$ are identical, since no more than one spin flip per site is possible.  However, for general spin-$s$ systems, up to $2s$ spin flips per site are possible, so that LSUB$n$ $\equiv$ SUB$2sn$-$n$.  Thus, for the spin-1 system, the SUB$n$-$n$ truncation contains fewer configurations than its LSUB$n$ counterpart for general values of the truncation index $n$.

Since the spin-$\frac{1}{2}$ $J_{1}$--$J_{2}$--$J_{1}^{\perp}$ model on a honeycomb-lattice bilayer in the limit $\delta \rightarrow -\infty$ clearly corresponds exactly to a spin-1 $J_{1}$--$J_{2}$ model on a honeycomb-lattice monolayer, the spin-1 monolayer result shown in Fig.\ \ref{M_multicontour_neel_honey_bilayer_AFM-FM} should provide a good limiting-case check of the present model.  Thus, we see that $\kappa_{c_{1}}^{>}(\delta \rightarrow -\infty)\approx 0.248$ from the spin-1 model using the SUB$n$-$n$ data set with $n=\{2,6,10\}$, which may be compared, for example, with the value $\kappa_{c_{1}}^{>}(\delta=-2.5) \approx 0.271$, as shown in Fig.\ \ref{M_multicontour_neel_honey_bilayer_AFM-FM}, and which is also obtained from the LSUB$n$ data set with $n=\{2,6,10\}$.  The closeness of the two values clearly demonstrates both the inherent accuracy of the CCM and the compatibility of the two different approximation schemes, LSUB$n$ and SUB$n$-$n$.

\begin{figure}[!t]
  \includegraphics[width=9cm]{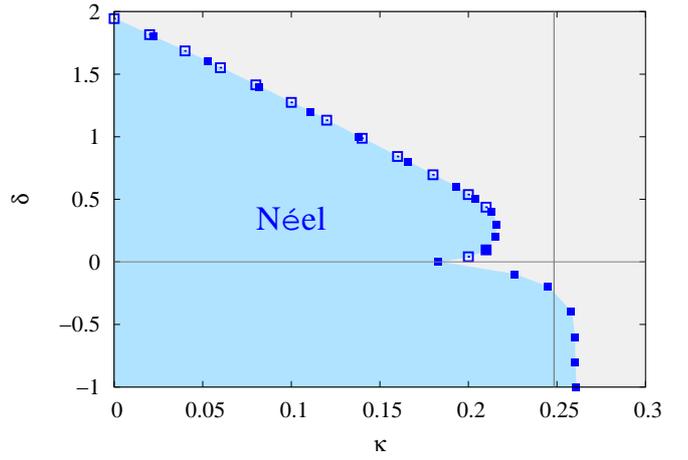}
  \caption{$T=0$ N\'{e}el phase diagram of the spin-$\frac{1}{2}$
    $J_{1}$--$J_{2}$--$J_{1}^{\perp}$ model on the bilayer honeycomb
    lattice with $J_{1}>0$, $\delta \equiv J_{1}^{\perp}/J_{1}$, and
    $\kappa \equiv J_{2}/J_{1}$.  The darker (blue) region is the quasiclassical GS phase with AFM N\'{e}el order in each monolayer (with the two layers coupled so that NN spins across them are aligned for $\delta<0$ and anti-aligned for $\delta > 0$), while in the lighter (grey) region N\'{e}el order is absent.  The filled and empty square symbols
    are points at which the extrapolated GS magnetic order parameter
    $M$ for the N\'{e}el phase vanishes, for specified values of
    $\delta$ and $\kappa$, respectively.  The N\'{e}el state on each
    monolayer is used as CCM model state, with the two layers coupled
    so that NN spins between them are parallel (antiparallel) to one another when $\delta<0\; (\delta>0)$, and
    Eq.\ (\ref{M_extrapo_frustrated}) is used for the extrapolations
    with the corresponding LSUB$n$ data sets $n=\{2,6,10\}$.  The vertical line at $\kappa=\kappa_{c_{1}}^{>}(\delta \rightarrow -\infty) \approx 0.248$ is taken from the $s=1$ honeycomb lattice monolayer result for $\kappa_{c_{1}}^{>}$, as explained in the text.}
\label{phase_diag_neel_honey_bilayer_AFM-FM}
\end{figure}

From curves such as those shown in Fig.\ \ref{M_multicontour_neel_honey_bilayer_AFM-FM}, from which we determined the values $\kappa_{c_{1}}^{>}(\delta)$ for various values of the scaled interlayer coupling parameter $\delta$, we can now plot the region in the $\kappa$-$\delta$ plane where the N\'{e}el state on each monolayer forms the stable GS phase, as shown in Fig.\ \ref{phase_diag_neel_honey_bilayer_AFM-FM}.
To the present results obtained for the case $\delta<0$, where the CCM model state has interlayer NN pairs aligned parallel to one another, we also display corresponding results obtained for the case $\delta > 0$, for which the CCM model state has the same pairs aligned antiparallel to one another \cite{Bishop:2017_honeycomb_bilayer_J1J2J1perp}.  Perhaps the most noteworthy feature of Fig.\ \ref{phase_diag_neel_honey_bilayer_AFM-FM} is the extremely sharp cusp in the phase diagram for the N\'{e}el phase, centered at $\delta=0$.  This feature clearly explains the great sensitivity that is observed in practice in most theoretical calculations of $\kappa_{c_{1}}^{>}(\delta=0)$, corresponding to the QCP for the vanishing of N\'{e}el order in the spin-$\frac{1}{2}$ $J_{1}$--$J_{2}$ model on a honeycomb-lattice monolayer.

\begin{figure}[!b]
  \includegraphics[width=9cm]{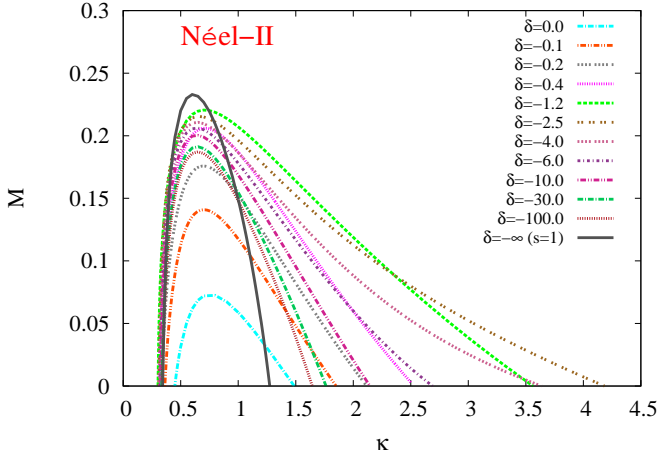}
  \caption{CCM results for the GS magnetic order parameter $M$ versus
    the intralayer frustration parameter, $\kappa \equiv J_{2}/J_{1}$,
    for the spin-$\frac{1}{2}$ $J_{1}$--$J_{2}$--$J_{1}^{\perp}$ model
    on the bilayer honeycomb lattice (with $J_{1}>0$), for a variety
    of ferromagnetic values of the scaled interlayer exchange coupling
    constant, $\delta \equiv J_{1}^{\perp}/J_{1}$.  In each case we
    show extrapolated results, based on the N\'{e}el-II state as CCM
    model state on each monolayer, and the two layers coupled so that
    NN spins between them are parallel to one another, obtained from
    using Eq.\ (\ref{M_extrapo_frustrated}) with the corresponding
    LSUB$n$ data sets $n=\{2,6,10\}$.  See the text for an explanation of the curve labeled $\delta=-\infty$ $(s=1$).}
\label{M_multicontour_neel2_honey_bilayer_AFM-FM}
\end{figure}

\begin{figure}[!b]
  \includegraphics[width=9cm]{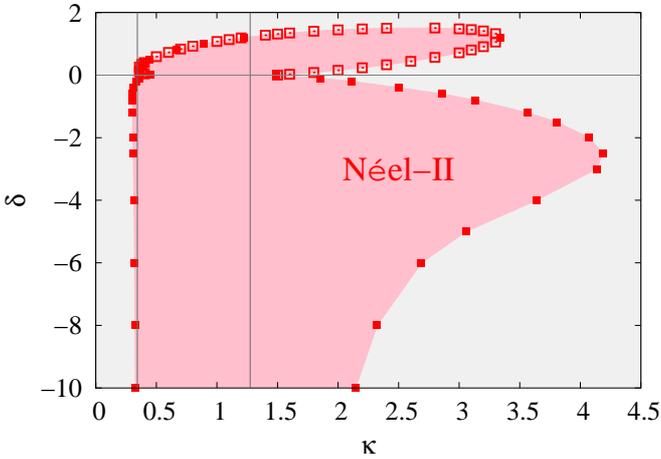}
  \caption{$T=0$ N\'{e}el-II phase diagram of the spin-$\frac{1}{2}$
    $J_{1}$--$J_{2}$--$J_{1}^{\perp}$ model on the bilayer honeycomb
    lattice with $J_{1}>0$, $\delta \equiv J_{1}^{\perp}/J_{1}$, and
    $\kappa \equiv J_{2}/J_{1}$.  The darker (pink) region is the quasiclassical GS phase with AFM N\'{e}el-II order in each monolayer (with the two layers coupled so that NN spins across them are aligned for $\delta<0$ and anti-aligned for $\delta > 0$), while in the lighter (grey) region N\'{e}el-II order is absent.  The filled and empty square symbols
    are points at which the extrapolated GS magnetic order parameter
    $M$ for the N\'{e}el phase vanishes, for specified values of
    $\delta$ and $\kappa$, respectively. The N\'{e}el-II state on each
    monolayer is used as CCM model state, with the two layers coupled
    so that NN spins between them are parallel (antiparallel) to one another when $\delta < 0$ ($\delta > 0$), and
    Eq.\ (\ref{M_extrapo_frustrated}) is used for the extrapolations
    with the corresponding LSUB$n$ data sets $n=\{2,6,10\}$.  The vertical lines at $\kappa = \kappa_{c_{2}}^{<}(\delta \rightarrow -\infty)\approx 0.343$ and $\kappa=\kappa_{c_{2}}^{>}(\delta = -\infty) \approx 1.274$ are taken from the $s=1$ honeycomb-lattice monolayer results for $\kappa_{c_{2}}^{<}$ and $\kappa_{c_{2}}^{>}$, respectively, as explained in the text. }
\label{phase_diag_neel2_honey_bilayer_AFM-FM}
\end{figure}

Similar to Fig.\ \ref{M_multicontour_neel_honey_bilayer_AFM-FM} for the N\'{e}el phase, we show in Fig.\ \ref{M_multicontour_neel2_honey_bilayer_AFM-FM} a set of LSUB$\infty$ extrapolated curves for the magnetic order parameter $M=M(\kappa)$ in the N\'{e}el-II GS phase, analogous to the one shown in Fig.\ \ref{M_neel_neel-II_fix-J1perp}(b) for the particular value $\delta=-1.2$.  We again display results for various values of $\delta(<0)$ in the regime where the two layers are coupled ferromagnetically (i.e., with NN interlayer pairs aligned parallel to each other.  Once again we plot the value $\mu_{0}$ obtained from fitting Eq.\ (\ref{M_extrapo_frustrated}) to the corresponding data set for $M(n)$ with $n=\{2,6,10\}$.  As in Fig.\ \ref{M_multicontour_neel_honey_bilayer_AFM-FM} for the N\'{e}el phase, we also shown in Fig.\ \ref{M_multicontour_neel2_honey_bilayer_AFM-FM} the corresponding CCM result for the N\'{e}el-II GS phase of the spin-1 $J_{1}$--$J_{2}$ model on a honeycomb-lattice monolayer, where we have again used Eq.\ (\ref{M_extrapo_frustrated}) to extract the extrapolated value $\mu_{0}$ from fits to the SUB$n$-$n$ data set with $n=\{2,6,10\}$.  In this case we have extended the results given in Ref.\ \cite{Li:2016_honeyJ1-J2_s1}, which were presented only for the range $0 \leq \kappa \leq 1$ there (and see Fig.\ \ref{phase_diag_neel_honey_bilayer_AFM-FM} of Ref.\ \cite{Li:2016_honeyJ1-J2_s1}), so as to also extract the value $\kappa_{c_{2}}^{>}$ for the model.  We find that, for this model, $\kappa_{c_{2}}^{<} \approx 0.343$ and $\kappa_{c_{2}}^{>}\approx 1.274$, which values should now also provide good estimates for the limiting values, $\kappa_{c_{2}}^{<}(\delta\rightarrow -\infty$) and $\kappa_{c_{2}}^{>}(\delta \rightarrow -\infty)$, of the current spin-$\frac{1}{2}$ $J_{1}$--$J_{2}$--$J_{1}^{\perp}$ honeycomb-lattice bilayer model.  For the purposes of comparison we quote the corresponding results, $\kappa_{c_{2}}^{<}(\delta=-30)\approx 0.335$ and $\kappa_{c_{2}}^{>}(\delta=-30)\approx 1.763$ for $\delta=-30$, and $\kappa_{c_{2}}^{<}(\delta=-100)\approx 0.337$ and $\kappa_{c_{2}}^{>}(\delta=-100) \approx 1.643$ for $\delta=-100$.  Clearly, the totally independent spin-1 monolayer results are in very good accord with the current spin-$\frac{1}{2}$ bilayer results.

Using curves such as those shown in Fig.\ \ref{M_multicontour_neel2_honey_bilayer_AFM-FM} to extract the values $\kappa_{c_{2}}^{<}(\delta)$ and $\kappa_{c_{2}}^{>}(\delta)$ for a variety of values of $\delta$, we can now also plot the region in the $\kappa$-$\delta$ plane where the N\'{e}el-II state on each monolayer forms the stable GS phase.  The result is shown in Fig.\ \ref{phase_diag_neel2_honey_bilayer_AFM-FM} where, in addition to the present results obtained for the case $\delta<0$, where the CCM model has interlayer NN spins aligned parallel to each other, we also display corresponding results obtained for the case $\delta>0$, for which the CCM model state has the same interlayer NN pairs aligned antiparallel to each other \cite{Li:2019_honeycomb_bilayer_J1J2J1perp_neel-II}.  Just as in Fig.\ \ref{M_multicontour_neel_honey_bilayer_AFM-FM} for the N\'{e}el phase, so the phase diagram in Fig.\ \ref{phase_diag_neel2_honey_bilayer_AFM-FM} for the N\'{e}el-II phase exhibits two sharp cusps centered at $\delta=0$.  These explain the sensitivity in providing accurate estimates for $\kappa_{c_{2}}^{<}(\delta=0)$ and $\kappa_{c_{2}}^{>}(\delta=0)$, which correspond to the QCPs for the disappearance of N\'{e}el-II AFM order in the spin-$\frac{1}{2}$ $J_{1}$--$J_{2}$ model on a honeycomb-lattice monolayer.  

\begin{figure}[!t]
  \includegraphics[width=9cm]{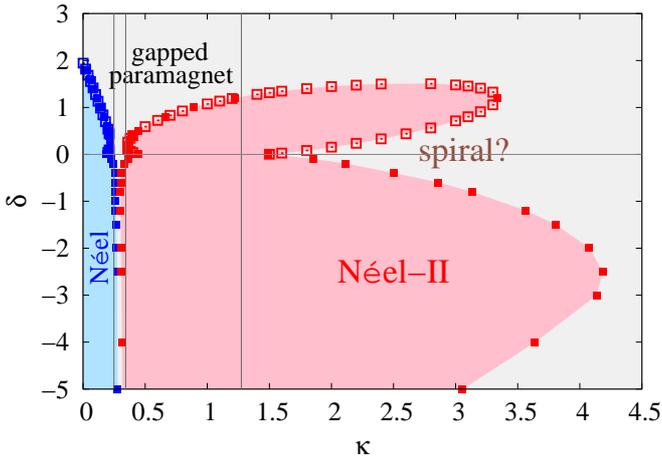}
  \caption{$T=0$ phase diagram of the
    spin-$\frac{1}{2}$ $J_{1}$--$J_{2}$--$J_{1}^{\perp}$ model on the
    bilayer honeycomb lattice with $J_{1}>0$,
    $\delta \equiv J_{1}^{\perp}/J_{1}$, and
    $\kappa \equiv J_{2}/J_{1}$.  The darker (blue and pink) shaded regions are the quasiclassical GS phases with AFM N\'{e}el and N\'{e}el-II orders in each monolayer, respectively (with the two layers coupled so that NN spins across them are aligned for $\delta<0$ and anti-aligned for $\delta>0$), while in the lighter (grey) shaded region quasiclassical collinear order is absent.   The filled and empty squares
    are points at which the extrapolated GS magnetic order parameter for the two quasicallical AFM phases vanishes, for specified values of
    $\delta$ and $\kappa$, respectively. In each case the N\'{e}el or N\'{e}el-II
    state on each layer is used as CCM model state, with the two
    layers coupled so that NN spins between them are parallel (antiparallel) to one
    another when $\delta<0$ ($\delta>0$), and Eq.\ (\ref{M_extrapo_frustrated}) is used for the
    extrapolations with the corresponding LSUB$n$ data sets
    $n=\{2,6,10\}$.  The vertical lines at $\kappa = \kappa_{c_{1}}^{>}(\delta = -\infty)\approx 0.248$, $\kappa = \kappa_{c_{2}}^{<}(\delta = -\infty)\approx 0.343$, and $\kappa=\kappa_{c_{2}}^{>}(\delta = -\infty) \approx 1.274$ are taken from the $s=1$ honeycomb-lattice monolayer results for $\kappa_{c_{1}}^{>}$, $\kappa_{c_{2}}^{<}$, and $\kappa_{c_{2}}^{>}$, respectively, as explained in the text.}
\label{phase_diag_neel_neel2_honey_bilayer_AFM-FM}
\end{figure}

Finally, in Fig.\ \ref{phase_diag_neel_neel2_honey_bilayer_AFM-FM} we combine all of our results to show the complete $T=0$ quasiclassical (N\'{e}el and N\'{e}el-II) phase diagram of our bilayer model.   The darker (blue and pink) shaded regions show, respectively, where the stable GS phase has AFM N\'{e}el and N\'{e}el-II order on each monolayer, while in the lighter (grey) shaded region all (quasiclassical) collinear magnetic order is absent in each monolayer.

\section{Discussion}
\label{discussion_sec}
We have implemented the powerful CCM technique of microscopic quantum many-body theory (QMBT) to very high orders in a systematic (LSUB$n$) hierarchy of approximations, to investigate the regions of stability of two quasiclassical collinear magnetically-ordered phases of a frustrated spin-$\frac{1}{2}$ $J_{1}$--$J_{2}$--$J_{1}^{\perp}$ Heisenberg magnet on a $AA$-stacked honeycomb-lattice bilayer, in the case where the intralayer NN and NNN couplings are both antiferromagnetic in nature (i.e., $J_{1}>0$ and $J_{2}>0$), for all values of the interlayer NN coupling parameter $J_{1}^{\perp}$.  The dual features of the CCM that it satisfies both the Goldstone linked cluster theorem and the Hellmann-Feynman theorem at all levels of approximation are unmatched by almost all alternative techniques of {\it ab initio\,} QMBT, and thus make it an ideal method for our purposes.  An immediate consequence of satisfying the former theorem is that we have been able to perform all calculations from the outset in the thermodynamic (infinite-lattice, $N \rightarrow \infty$) limit.  As a result we have never had to resort to any finite-size scaling of our numerical results, as is required in almost all other high-accuracy techniques, and which is then often the major cause of uncertainty in their results.

Our sole approximation has thus been to extrapolate numerically the sequences of LSUB$n$ approximants for the GS energy and GS magnetic order parameter to the $n \rightarrow \infty$ limit, where the CCM becomes exact.  By implementing the method computationally with the purpose-built computer-algebra (CCCM) package \cite{ccm_code} to very high orders (viz., with $n \leq 10$), we have been able to perform the extrapolations with very high accuracy.  In order to do so we have circumvented not only the well-known $(2m-1)/2m$ staggering effect (where $m \in \mathbb{Z}^{+}$ is a positive integer) in the sequences of $n$th order approximants for various physical observables, which is always present, but also the additional (and less well known) $(4m-2)/4m$ staggering effect of the even $(n=2m)$ subsequences, which the honeycomb lattice also exhibits as a probable consequence of its non-Bravais nature, by performing all LSUB$n$ extrapolations with the limited data sets $n=\{2,6,10\}$.

Given that we have fitted our LSUB$n$ results for the magnetic order parameter $M$ with the data set $n=\{2,6,10\}$ to the three-parameter fit of Eq.\ (\ref{M_extrapo_frustrated}), it is not possible in this case to give a precise estimate of any systematic errors in our results.  Nevertheless, many prior applications of the methodology have shown that the extrapolation scheme is very robust.  In general, in cases where we have more LSUB$n$ data points than three, a least-squares fit gives a reliable estimate of the errors.  Many such estimates have been given in prior publications involving the CCM, particularly for applications on comparable models on Bravais lattices for which only the $(2m-1)/2m$ staggering effect is present, and in which the $(4m-2)/4m$ staggering effect is absent, and hence for which more LSUB$n$ data points are available to perform the extrapolations.  A good example can be found in Ref.\ \cite{Bishop:2019_SqLatt_bilayer} where results have been presented for the analogous model to that presented here, but on a square-lattice bilayer.  The high accuracy of the extrapolations demonstrated explicitly in that case gives us considerable confidence in the location of the phase boundaries in our final phase diagram of Fig.\ \ref{phase_diag_neel_neel2_honey_bilayer_AFM-FM}.

We note that for the case of the honeycomb-lattice monolayer (i.e., with $\delta=0$), N\'{e}el-II order does not exist classically as the stable GS of the $J_{1}$--$J_{2}$ model, except precisely at the point $\kappa=\frac{1}{2}$, which is a point of maximal GS degeneracy.  Thus, for the classical $J_{1}$--$J_{2}$ model on the honeycomb lattice (with $J_{1}>0$, $J_{2}>0$) N\'{e}el order exists for all values $\kappa \leq \frac{1}{6}$.  Conversely, for all values $\kappa > \frac{1}{6}$ the model has an infinitely degenerate family of incommensurate, coplanar, spiral configurations of spins, described by an ordering wave vector $\mathbf{Q}$, for which the direction is arbitrary.  For values of $\kappa$ in the range $\frac{1}{6} < \kappa < \frac{1}{2}$ these classically degenerate solutions for $\mathbf{Q}$ form a closed contour around the center point $\mathbf{Q} = \mathbf{\Gamma} \equiv (0,0)$ of the hexagonal first Brillouin zone.  By contrast, when $\kappa > \frac{1}{2}$, the solutions for $\mathbf{Q}$ lie on pairs of closed contours that are now centered on any two of the inequivalent corners of the first Brillouin zone.  The point $\kappa = \frac{1}{2}$ marks a second classical transition point between two different forms of GS spiral phases, and it is precisely at this point that both are also degenerate with the collinear N\'{e}el-II phase.

We have seen from Fig.\ \ref{M_multicontour_neel2_honey_bilayer_AFM-FM} that for the case $\delta=0$ of the spin-$\frac{1}{2}$ $J_{1}$--$J_{2}$--$J_{1}^{\perp}$ model on the bilayer honeycomb lattice, quantum fluctuations have now stabilized N\'{e}el-II order over the range $\kappa_{c_{2}}^{<}(\delta=0) \approx 0.45 < \kappa < \kappa_{c_{2}}^{>}(\delta=0) \approx 1.49$, in keeping with the general observation that quantum fluctuations tend to favor collinear over non-collinear forms of order.  Nevertheless, the N\'{e}el-II order present is quite fragile, as may be seen from the rather small values of the order parameter $M$ in the $\delta=0$ curve in Fig.\ \ref{M_multicontour_neel2_honey_bilayer_AFM-FM} over the entire region of its existence (i.e., when $M > 0$).  By contrast, Fig.\ \ref{M_multicontour_neel2_honey_bilayer_AFM-FM} also shows that even a small interlayer coupling enhances the stability of the N\'{e}el-II phase considerably.

Although it is out of the scope of the present work to study in detail the nature of the GS phases of the model in the lighter (grey) shaded areas of Fig.\ \ref{phase_diag_neel_neel2_honey_bilayer_AFM-FM}, where collinear magnetic ordering is absent, nevertheless it is interesting to speculate.  Thus, for example, from our discussion above, we are led to expect that in the region around the $\delta=0$ axis, and between the two lobes of the region of stability where N\'{e}el-II ordering is present on each monolayer, the GS phase will have spiral ordering.  This assertion has been lent credence, again by the use of the CCM, but now to study excited states also, by the finding \cite{Li:2019_honeycomb_bilayer_J1J2J1perp_neel-II} that (at least part of) this region is gapless.  From the same CCM study of the excitation energy to the lowest excited state of the model, the clear presence of a gapped state has also been shown to exist over much of the paramagnetic region in Fig.\ \ref{phase_diag_neel_neel2_honey_bilayer_AFM-FM} between the two islands of stability (shown by the darker, blue or, pink, shaded regions) in which N\'{e}el or N\'{e}el-II magnetic ordering is present on each monolayer.  It is likely that this gapped paramagnetic region comprises one or more types of VBC phase, including those of the staggered-dimer (SDVBC) or hexagonal plaquette (PVBC) variety on each monolayer.  Another type, namely the interlayer dimer (IDVBC) variety, also seems bound to occur, particularly for larger absolute values $|\delta|$ of the interlayer coupling, where the dimers now occur between NN pairs of spins across the two $AA$-stacked layers.

In practical applications of the current model, the bilayer may well rest on top of a suitable substrate.  One way to simulate the effects of such a substrate would be to include disorder on the system.  It would then be of considerable interest to investigate which phases and, particularly, which phase boundaries, in Fig.\ \ref{phase_diag_neel_neel2_honey_bilayer_AFM-FM} are robust against weak on-site potential disorder.

\section*{ACKNOWLEDGMENTS}
We thank the University of Minnesota Supercomputing Institute for the
grant of supercomputing facilities, on which some of the work reported here
was performed.  The authors would also like to acknowledge the assistance given by Research IT and the use of the HPC Pool funded by the Research Lifecycle Programme at the University of Manchester.  One of us (RFB) gratefully acknowledges the Leverhulme Trust (United Kingdom) for the award of an Emeritus Fellowship (Grant No. EM-2020-013).

\section*{References}



\bibliographystyle{elsarticle-num} 
\bibliography{bib_general}





\end{document}